\documentclass[acmtog]{acmart}
\acmSubmissionID{319}

\usepackage{booktabs} 
\usepackage{wrapfig}
\usepackage{subcaption}
\usepackage{siunitx} 

\usepackage[ruled]{algorithm2e} 

\SetAlFnt{\small}
\SetAlCapFnt{\small}
\SetAlCapNameFnt{\small}
\SetAlCapHSkip{0pt}

\setcopyright{acmlicensed}
\acmJournal{TOG}
\acmYear{2024} \acmVolume{43} \acmNumber{4} \acmArticle{69} \acmMonth{7}\acmDOI{10.1145/3658152}

\usepackage{xspace}
\makeatletter
\DeclareRobustCommand\onedot{\futurelet\@let@token\@onedot}
\def\@onedot{\ifx\@let@token.\else.\null\fi\xspace}
\def\eg{\emph{e.g}\onedot} 
\def\ie{\emph{i.e}\onedot}

\def\etal{\emph{et al}\onedot}
\makeatother

\usepackage[makeroom]{cancel}

\newcommand{\bc}{\mathbf{c}}

\newcommand{\bn}{\mathbf{n}}
\newcommand{\bp}{\mathbf{p}}

\newcommand{\bu}{\mathbf{u}}
\newcommand{\bv}{\mathbf{v}}
\newcommand{\bx}{\mathbf{x}}

\newcommand{\by}{\mathbf{y}}
\newcommand{\bZero}{\mathbf{0}}

\newcommand{\dd}{\mathrm{d}}

\DeclareMathOperator*{\argmin}{argmin}

\newif\ifcomment
\commenttrue

\begin{document}

\title{Differentiable Voronoi Diagrams for Simulation of Cell-Based Mechanical Systems}

\author{Logan Numerow}
\affiliation{%
  \institution{ETH Z{\"u}rich}
  \country{Switzerland}
}
\email{lnumerow@student.ethz.ch}

\author{Yue Li}
\affiliation{%
  \institution{ETH Z{\"u}rich}
  \country{Switzerland}
}
\email{yue.li@inf.ethz.ch}

\author{Stelian Coros}
\affiliation{%
  \institution{ETH Z{\"u}rich}
  \country{Switzerland}
}
\email{stelian.coros@inf.ethz.ch}

\author{Bernhard Thomaszewski}
\affiliation{%
  \institution{ETH Z{\"u}rich}
  \country{Switzerland}
}
\email{bthomasz@ethz.ch}

\begin{abstract}
Navigating topological transitions in cellular mechanical systems is a significant challenge for existing simulation methods. While abstract models lack predictive capabilities at the cellular level, explicit network representations struggle with topology changes, and per-cell representations are computationally too demanding for large-scale simulations.
To address these challenges, we propose a novel cell-centered approach based on differentiable Voronoi diagrams. Representing each cell with a Voronoi site, our method defines shape and topology of the interface network implicitly. 
In this way, we substantially reduce the number of problem variables, eliminate the need for explicit contact handling, and ensure continuous geometry changes during topological transitions. Closed-form derivatives of network positions facilitate simulation with Newton-type methods for a wide range of per-cell energies. Finally, we extend our differentiable Voronoi diagrams to enable coupling with arbitrary rigid and deformable boundaries.
We apply our approach to a diverse set of examples, 
highlighting splitting and merging of cells as well as neighborhood changes. We illustrate applications to inverse problems by matching soap foam simulations to real-world images. Comparative analysis with explicit cell models reveals that our method achieves qualitatively comparable results at significantly faster computation times.
\end{abstract}

%



\begin{teaserfigure}
   \centering
    \begin{subfigure}{.24\linewidth}
      \centering
      \includegraphics[width=\textwidth]{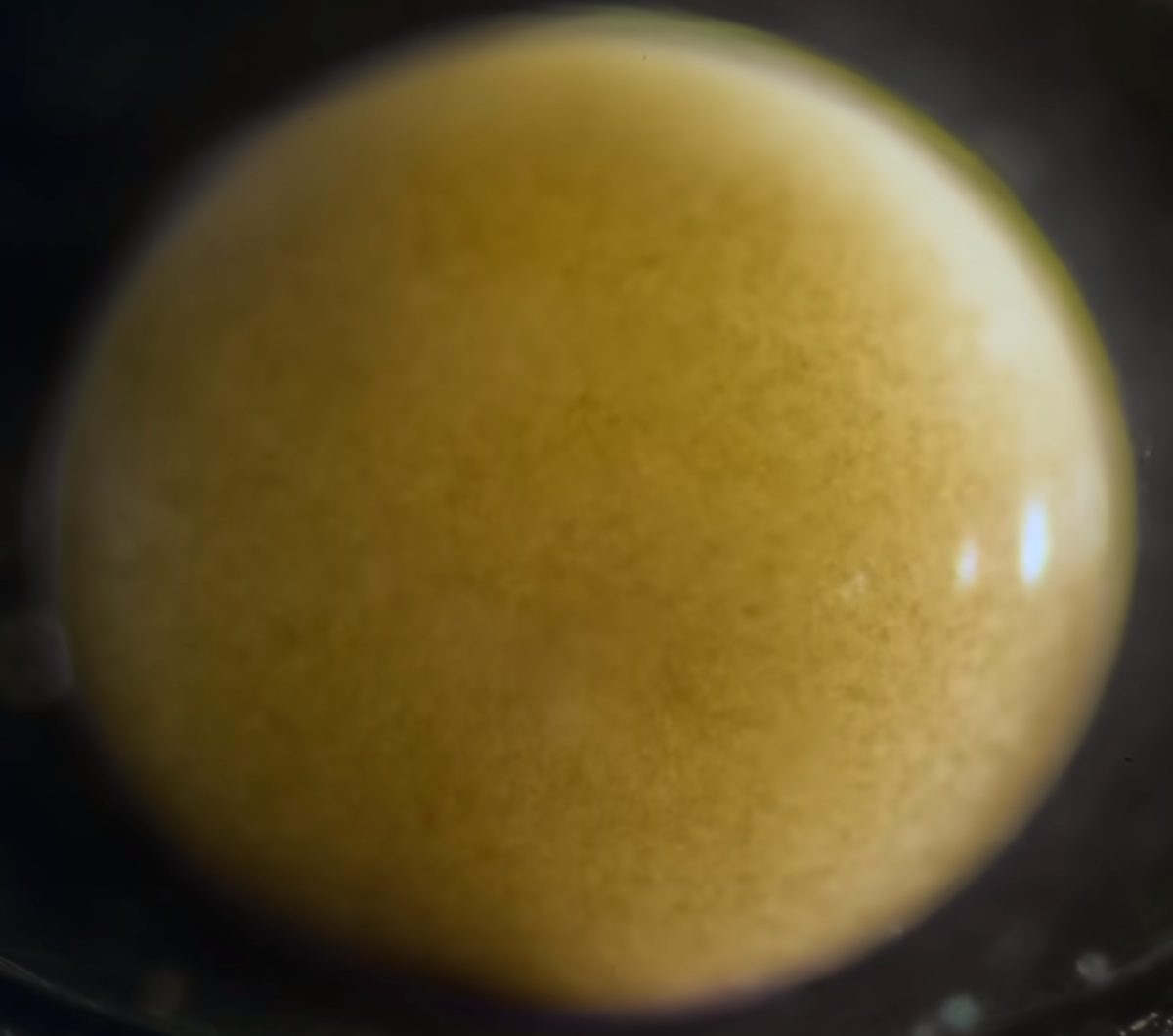}
    \end{subfigure}
    \begin{subfigure}{.24\linewidth}
      \centering
      \includegraphics[width=\textwidth]{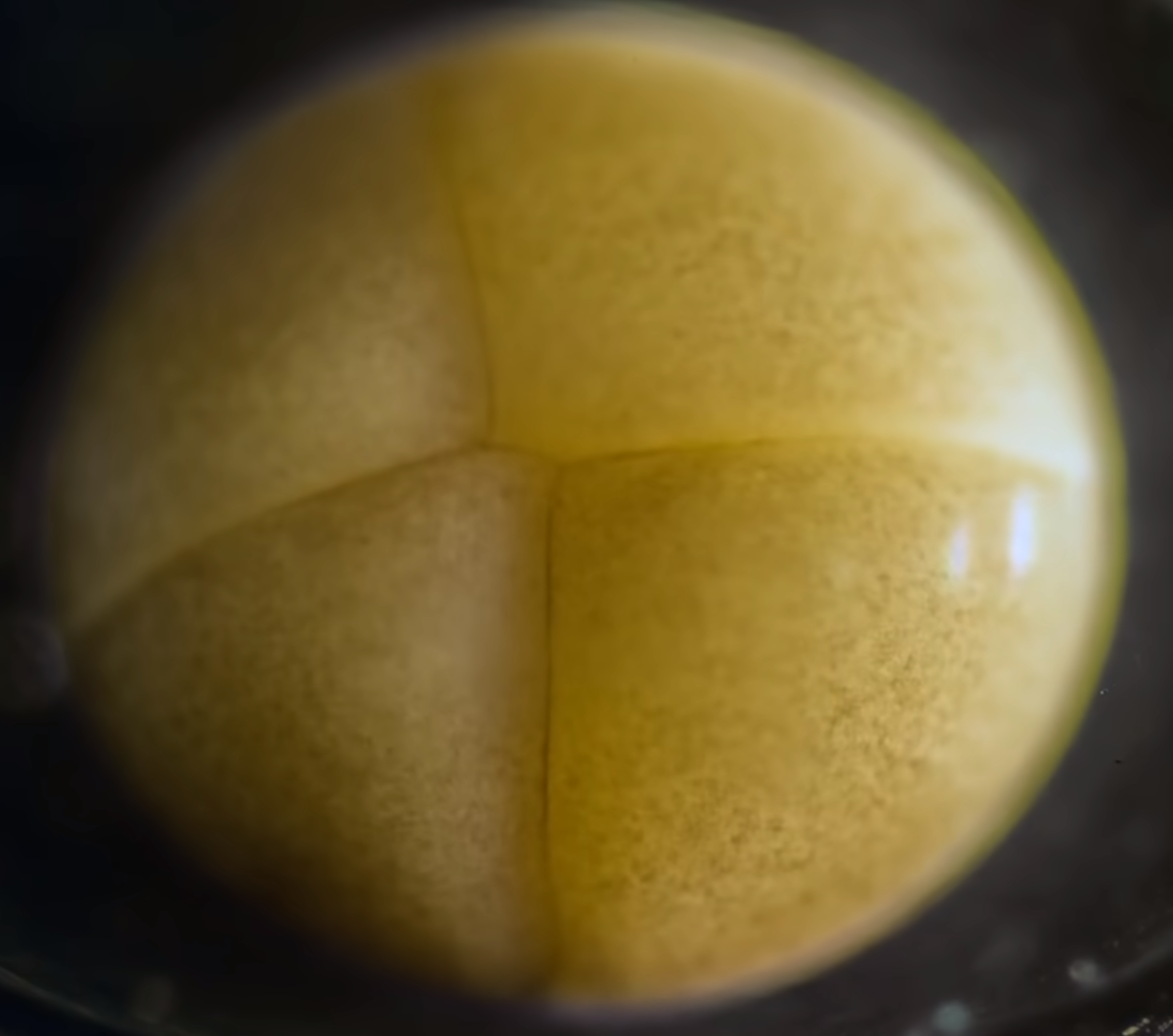}
    \end{subfigure}
    \begin{subfigure}{.24\linewidth}
      \centering
      \includegraphics[width=\textwidth]{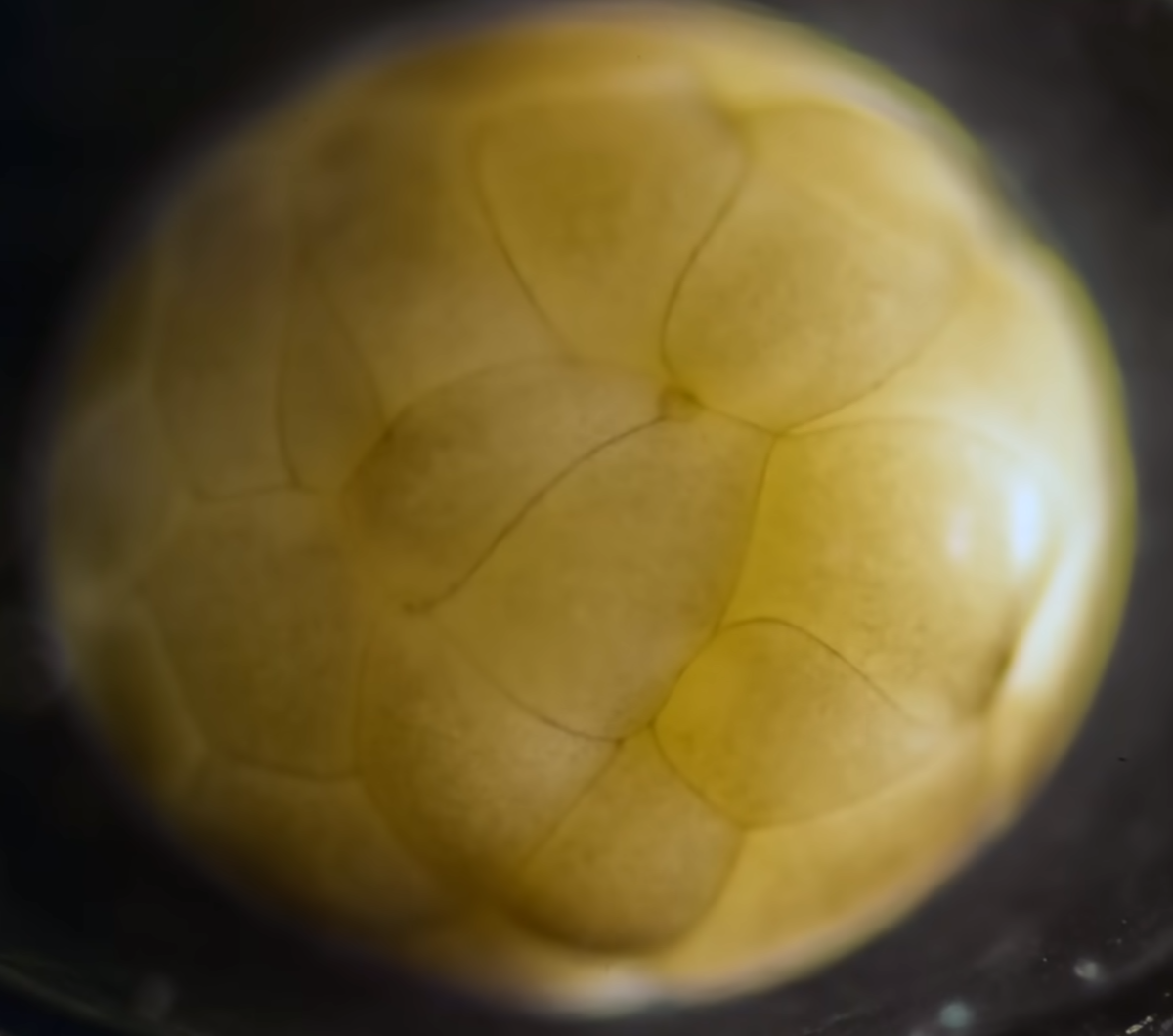}
    \end{subfigure}
    \begin{subfigure}{.24\linewidth}
      \centering
      \includegraphics[width=\textwidth]{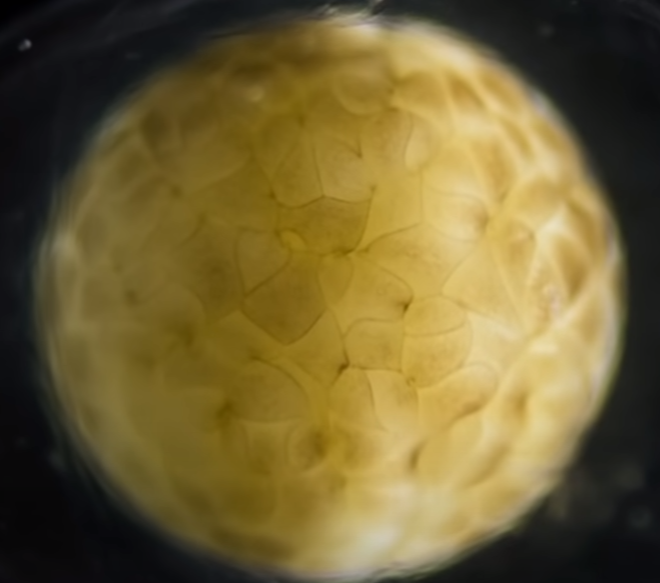}
    \end{subfigure}
    \begin{subfigure}{.24\linewidth}
      \centering
      \includegraphics[width=0.9\textwidth]{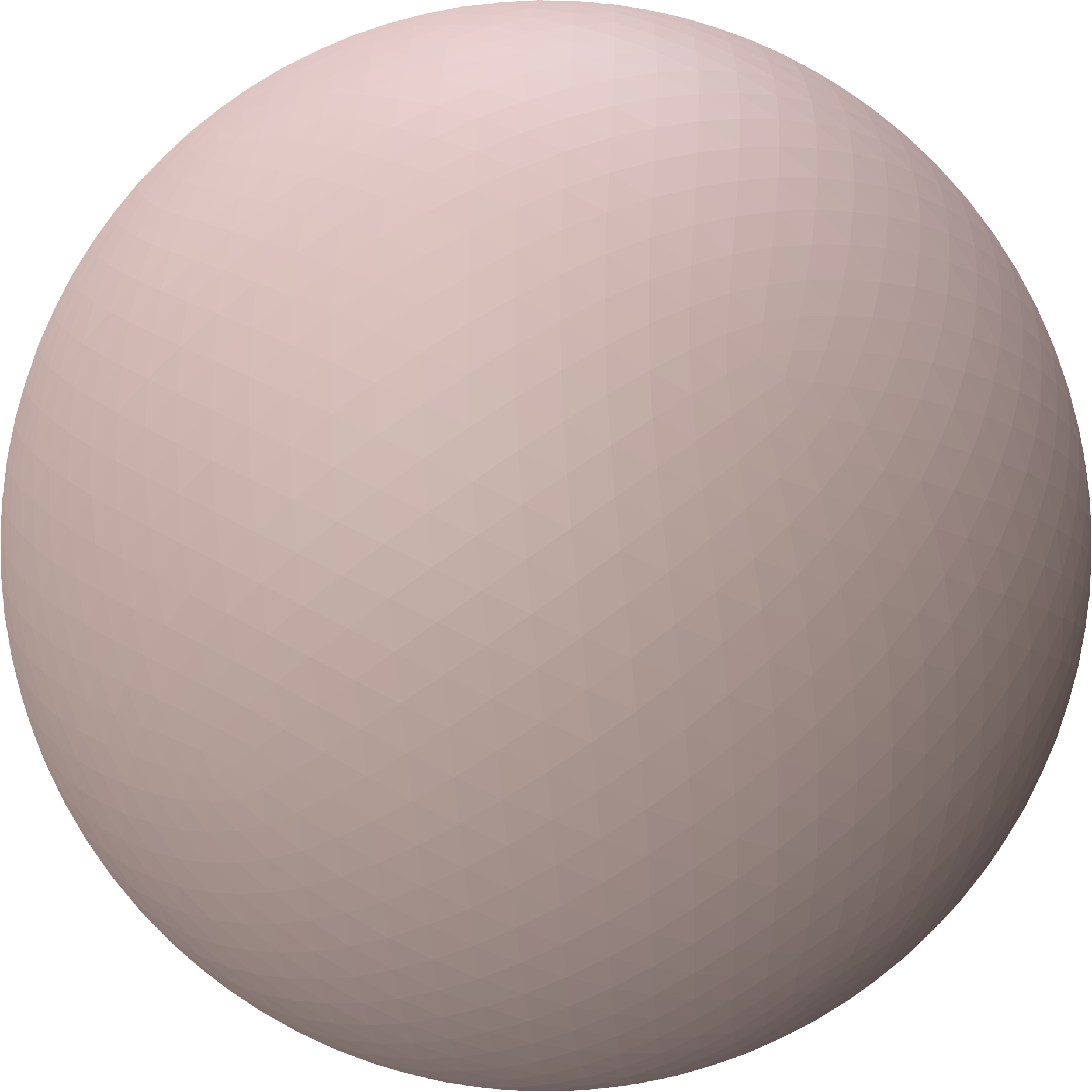}
    \end{subfigure}
    \begin{subfigure}{.24\linewidth}
      \centering
      \includegraphics[width=0.9\textwidth]{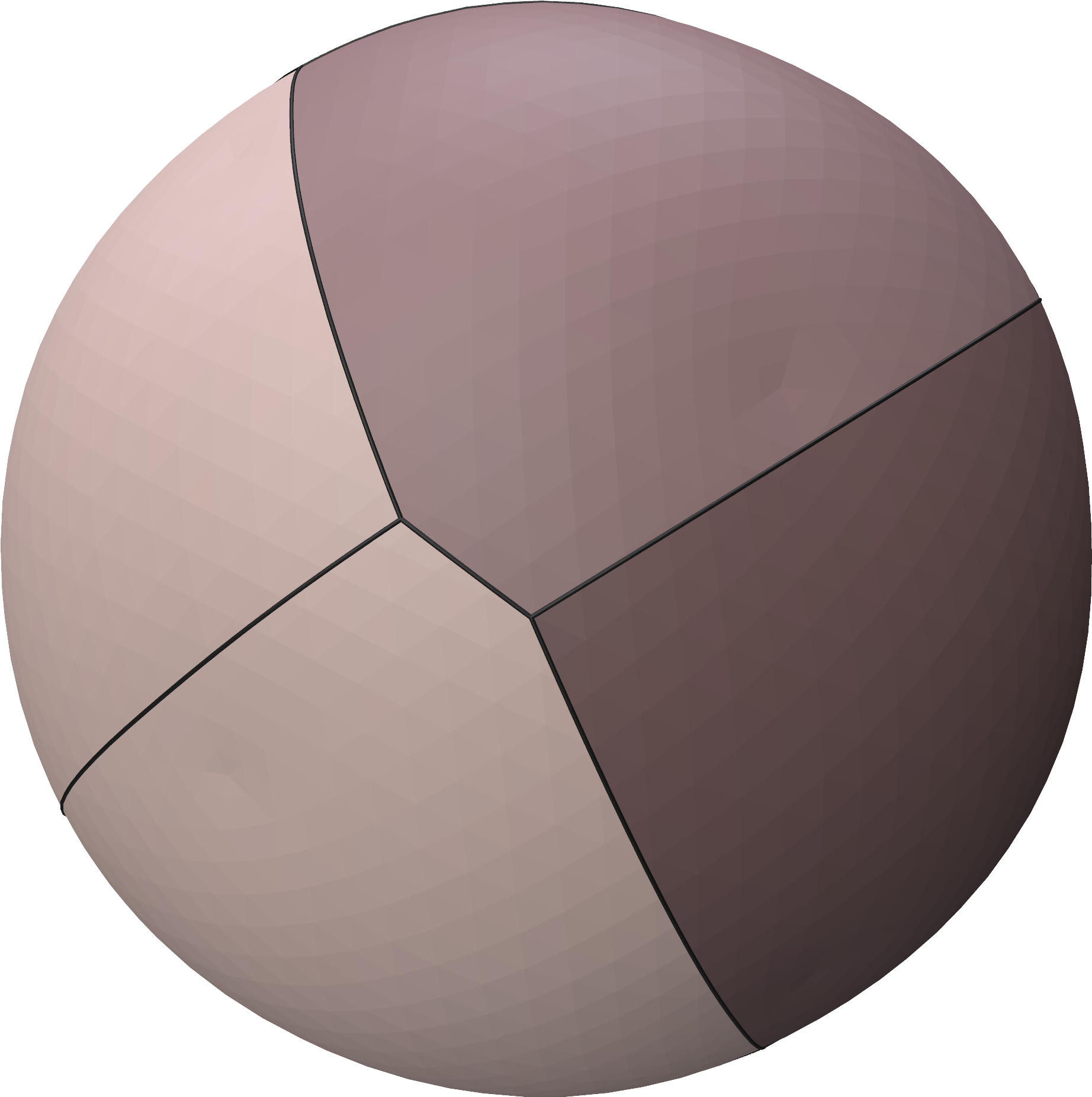}
    \end{subfigure}
    \begin{subfigure}{.24\linewidth}
      \centering
      \includegraphics[width=0.9\textwidth]{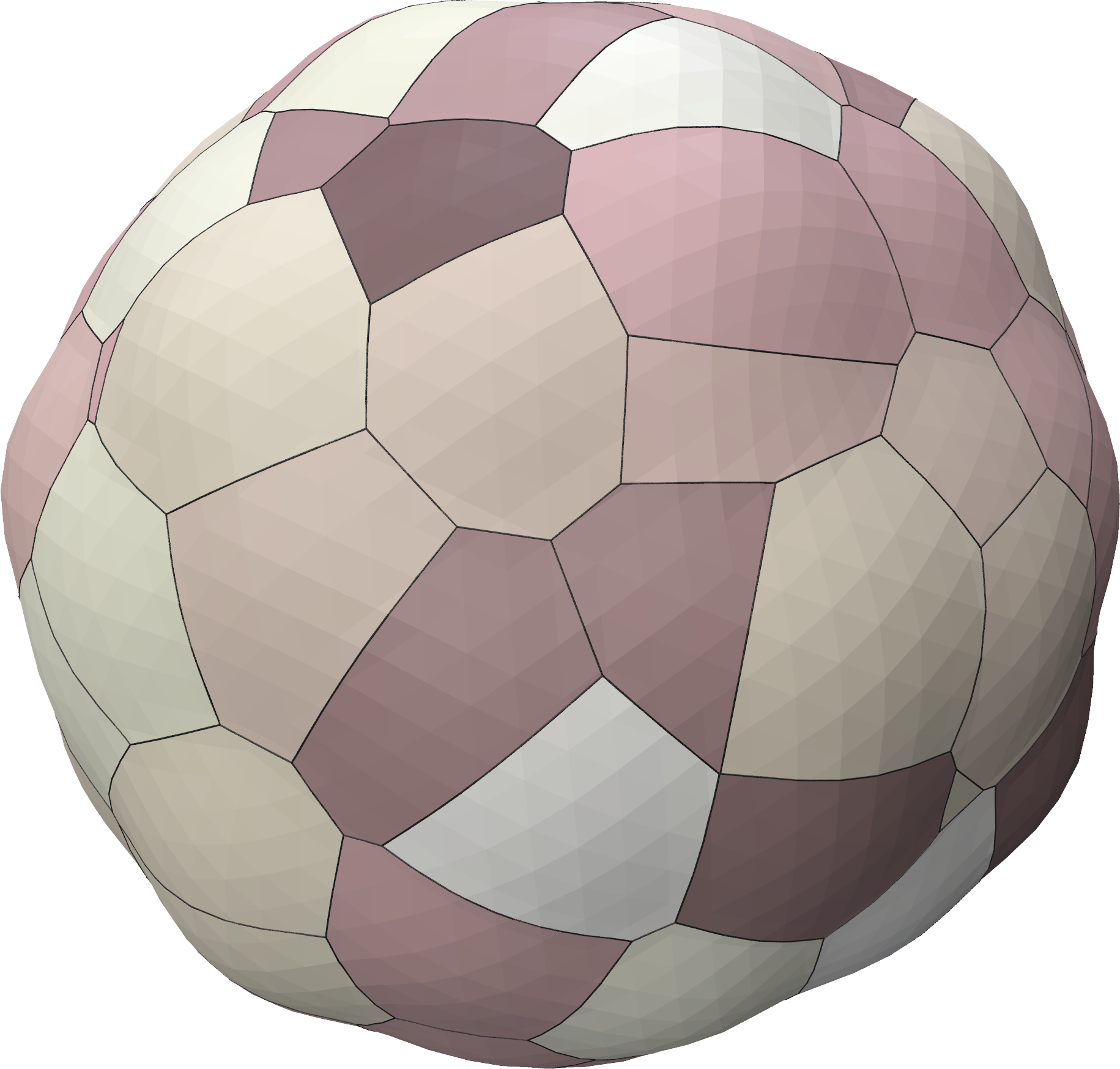}
    \end{subfigure}
    \begin{subfigure}{.24\linewidth}
      \centering
      \includegraphics[width=0.9\textwidth]{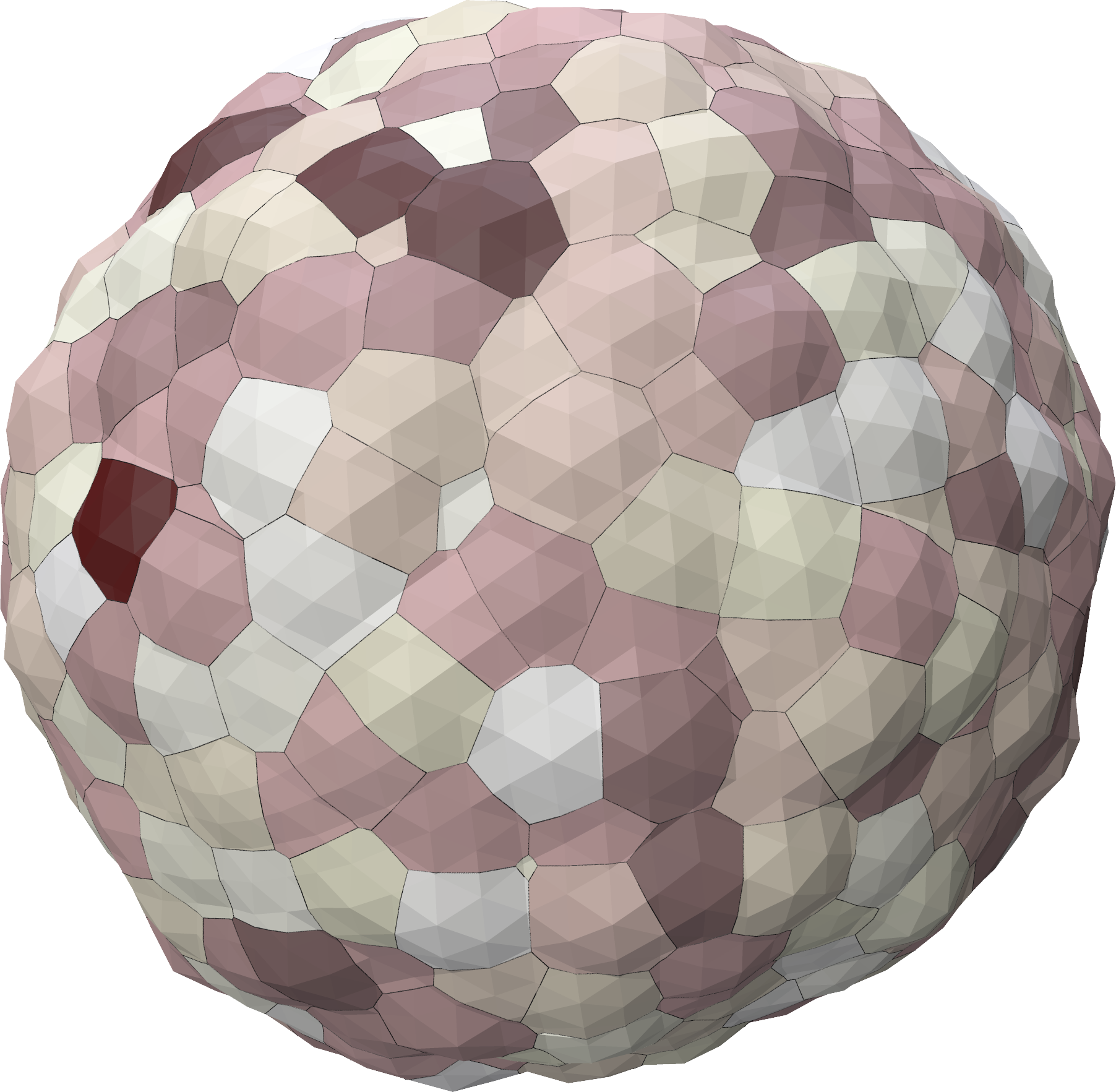}
    \end{subfigure}
   \caption{We use differentiable Voronoi diagrams to simulate embryonic cleavage in an elastic spherical membrane. \textit{Bottom}: As cells split according to a predefined schedule, substantial neighborhood changes and cell deformations occur. Each cell is modeled using only a single Voronoi site carrying four degrees of freedom. \textit{Top}: Images from Jan van Ijken's "Becoming"~\shortcite{vanIjken}, depicting the embryonic development of a salamander, are included for comparison.}
\label{fig:cleavage_membrane}
\end{teaserfigure}
\maketitle

\section{Introduction}

A central characteristic of cellular mechanical systems is that cells change neighborhoods during dynamic motion and growth-induced splitting (as in biological tissue), or through merging and collapse (as in soap foams). These topological transitions cause abrupt force changes, which make computational modeling very challenging.

There are several existing approaches in the literature for simulation of cell-based systems. One option is to abstract away the cellular structure and model the large-scale mechanics of the system in a homogenized sense. Evidently, such models cannot predict interactions at the cellular level governed by local topology changes.
Another approach is to explicitly model the network of cell interfaces, enabling efficient computation of cell volumes and other geometric quantities that influence their mechanics. However, topology changes are difficult to model with explicit network representations, requiring discrete mesh adaptations and other discontinuous operations.
Another class of models represents each cell individually using, \eg, per-cell finite element discretizations. While this approach allows for accurate modeling of cell interactions, it requires a very large number of degrees of freedom to resolve cell deformations, and contact must be handled explicitly at every cell interface. These aspects render this strategy unsuitable for large-scale simulations involving hundreds or thousands of cells.

To overcome the limitations of existing methods, we propose a novel cell-centered approach based on generalized Voronoi diagrams. Each cell is represented by a Voronoi \textit{site},  and the topology and shape of the cells are defined implicitly. This representation uses far fewer degrees of freedom, avoids explicit handling of contact between cells, and ensures continuous cell geometry changes through topological transitions.  
In order to use this Voronoi model for simulation, we derive closed-form expressions for first and second derivatives of cell geometry with respect to site positions, allowing us to use Newton-type optimization solvers for a wide range of energies. We furthermore obtain equilibrium-state derivatives of the interface network using sensitivity analysis, thus opening the door to a wide range of inverse simulation problems.
Finally, we extend our differentiable Voronoi formulation to enable coupling with arbitrary rigid and deformable boundaries.

We demonstrate the potential of our method on a diverse set of examples, including tissue growth and foam coarsening. We furthermore show applications of differentiable Voronoi diagrams to inverse problems by optimizing per-cell pressures to match input images from real-world foams.
Our comparison with explicit cell models indicates that differentiable Voronoi diagrams can produce qualitatively similar behavior with much faster computation times.

\section{Related Work}
\paragraph{Simulating Cell \& Tissue Dynamics}
Various computational models have been developed to simulate cell-based mechanical systems. One approach is to abstract away the cellular structure and model the mechanical properties of the system in an averaged sense at coarser scales
~\cite{Brodland2006, Kondo2018, Piatnitski2020}. Our focus, however, is on problems that are governed by cell-level phenomena.
One option in this context is to explicitly model the network of cell interfaces using so-called \textit{vertex models}~\cite{Honda2004, Fletcher2014, Alt2017}. Although extensions have been developed to allow for topology changes~\cite{Okuda2012, Farhadifar2007}, the discrete mesh modifications involved in these methods are problematic for differentiable simulation. Moreover, explicit vertex models require additional constraints to prevent self-intersections and other invalid geometry~\cite{VedelLarsen2010}. Our differentiable Voronoi diagrams, in contrast, ensure a valid interface network at all times and guarantee continuous topology changes for any displacement of the Voronoi sites.
\par
Another line of work has explored so-called \textit{deformable cell models}, in which cells are represented through separate surface meshes~\cite{Kim2021, PalaCell2021}. While this approach enables accurate modeling of contact forces and topology changes, it requires explicit handling of collisions between cells. Moreover, hundreds to thousands of degrees of freedom are needed per cell, which rapidly leads to all but intractable problem size. 
With each cell encoded by the position of a single Voronoi site, our approach requires dramatically fewer degrees of freedom. The resulting model is highly scalable, enabling efficient gradient-based optimization with thousands of cells.

\paragraph{Simulating Soap Films, Foams, \& Bubbles}
The dynamics of bubbles have attracted substantial interest from the graphics community. Examples include the formation of bubbles in water \cite{Hong08Bubbles}, soap film dynamics \cite{Ishida20SoapFilm,Huang20Chemomechanical,Deng22AMoving}, and general foams \cite{Busyarev2012}.
Various representations have been explored, ranging from explicit surface meshes \cite{Da15DoubleBubbles} to mixed particle-and-grid discretizations \cite{Goldade20Constraint}.
We show that, using only a single site per foam cell, differentiable Voronoi diagrams can produce realistic 3D foam coarsening simulations. Our method can likewise be used for inverse problems such as matching soap film simulations to real-world images.  
\paragraph{Implicit Voronoi Diagrams}
Voronoi diagrams have numerous applications in computer graphics, vision, and beyond. They have been explored as a basis for, \eg, fluid \cite{deGoes2015} and foam \cite{Busyarev2012} simulations, image segmentation~\cite{williams2020voronoinet} as well as for structural design. 
Zhang \etal~\shortcite{Zhang2017} design personalized medical braces represented as centroidal Voronoi diagrams. Their algorithm uses a variant of Lloyd’s method to iteratively adjust site positions in order to minimize the design objective.
Lumpe \etal\shortcite{Lumpe2023} investigate power diagrams for topology optimization. While cell volumes are adapted in each iteration, all simulations are performed on finite element meshes generated from the power diagrams.  
Also targeting topology optimization, Feng \etal\shortcite{Feng2022} introduce a smooth relaxation of Voronoi diagrams based on \textit{softmax} functions. However, their approach ultimately requires rasterization on a Cartesian grid for optimization.
\par
We refer to the aforementioned Voronoi models as \textit{implicit Voronoi diagrams}, as the simulated structures are defined implicitly by Voronoi sites rather than with explicit degrees of freedom.
Lloyd's algorithm \shortcite{Lloyd82} for generating centroidal Voronoi tessellations (CVT) is arguably the most widely used method for optimization of Voronoi diagrams.
Nevertheless, substantial accelerations can be achieved when using gradient-based or quasi-Newton methods \cite{CVT2009, Yan2009, Yan2011,wang2015intrinsic}. Unlike these methods, our work provides explicit derivatives of Voronoi vertices, opening the door to general energy functions that can be adapted to a large variety of physical systems.
In addition, we provide an algebraic recipe for unified treatment of Voronoi cell faces and clipping geometry, enabling simulation of cellular systems coupled to moving or deformable boundaries.
\paragraph{Adaptive Meshes}
Adapting meshes to ensure given quality criteria is a common strategy in many simulation contexts, including large elastoplastic deformations \cite{Wicke10,Bargteil07,wojtan2009deforming}, fracture \cite{OBrien02Graphical,Pfaff14,koschier2014adaptive}, fluid flow \cite{Misztal14Multiphase}, and the folding of thin sheet materials \cite{Narain12Adaptive,Narain13Folding,Schreck15Nonsmooth,Zhang22Progressive}. 
Remeshing is typically applied as a post-process after each time step, which can disturb force equilibrium and lead to abrupt configuration changes. Coupling remeshing with time integration can avoid such artifacts \cite{Ferguson23InTimestep}.
While mesh adaptation is often an optional improvement, topological changes that occur, e.g., during droplet formation make remeshing compulsory \cite{Brochu09}. The applications that we consider in this work likewise involve inherent topology changes as cells split, collapse, or switch neighbors. However, our differentiable Voronoi diagrams eliminate the need for explicit remeshing,  as the mesh is given implicitly through the Voronoi sites.
\paragraph{Equilibrium-Constrained Optimization}
Differentiable simulation is a core technology for solving equilibrium-constrained optimization problems in graphics and beyond. Examples include material parameter estimation \cite{Miguel12DataDriven,hahn2019real2sim,yan2018inexact,Bickel09Capture}, elastic shape optimization \cite{Chen14ANS,Ly18Inverse,Panetta17Worst}, inflatables \cite{Skouras14Designing,Panetta21Inflatables}, clothing design \cite{Umetani11Sensitive,Wang18RuleFree,Montes20Computational}, and trajectory optimization \cite{geilinger2020add,hu2019chainqueen}. 
Unlike existing methods that impose equilibrium constraints on explicit triangle or tetrahedron meshes, our formulation enables the solution of inverse problems defined on Voronoi diagrams.

\section{Differentiable Restricted Voronoi Diagrams}
\subsection{Generalized Voronoi Diagrams}
The Voronoi cell $\mathcal{R}_i$ for a given site $C_i$ can be defined as
\begin{equation}\label{eq:voronoi}
\mathcal{R}_i = \{p : d(p, C_i) < d(p, C_j) \forall j \neq i\}\ ,
\end{equation}
where $d(p,C) = d_{\text{Euclidean}}(p,C)$ is the Euclidean distance between point $p$ and site $C$.
A family of more flexible generalized Voronoi diagrams can be constructed by modifying the distance metric which defines the Voronoi cell. 
\par
Mouzarkel~\shortcite{Moukarzel1997} and Eppstein~\shortcite{Eppstein2012} draw theoretical links between foam structures and generalized Voronoi diagrams, independently showing that a two-dimensional foam at equilibrium is geometrically equivalent to a \textit{sectional multiplicative Voronoi partition} (SMVP),
\begin{equation}\label{eq:smvp}
d_{\text{SMVP}}(p,C_i)^2 = \frac{1}{k_i}\left(d_{\text{Euclidean}}(p,C_i)^2 - w_i\right) \ ,
\end{equation}
for some arrangement of sites and weights $w$, $k$. The multiplicative weight $k$ gives rise to curved interfaces which, while necessary to represent exact foam structures, significantly complicate computations of geometric properties of cells (Sec.~\ref{sec:cell_integration}). Analytic integration over 3D curved surfaces such as soap films is typically impossible, and algebraic line-surface intersection computations, required for boundary clipping, involve solving high-order polynomials even for simple 3D interpolating surfaces. We instead use the power diagram,
\begin{equation}\label{eq:power}
d_{\text{Power}}(p,C_i)^2 = d_{\text{Euclidean}}(p,C_i)^2 - w_i \ ,
\end{equation}
for our simulations, which maintains planar interfaces while allowing for cell size variability. Sec.~\ref{sec:results_image_matching} demonstrates the power diagram's capacity to express realistic foam geometry.
%
\subsection{Voronoi Vertices}
We compute the restricted Voronoi diagram, constrained to within a bounding domain, as described by Yan \etal~\shortcite{Yan2009, Yan2011}. Each restricted Voronoi vertex in $m$ dimensions is an intersection of $m$ hyperplanes, each of which is either a Voronoi bisector or a facet of the bounding surface. In three dimensions, there are four types of vertices; (a) \textit{unrestricted Voronoi vertices}, \ie, intersections between three
\setlength{\columnsep}{3pt}
\begin{wrapfigure}{r}{0.25\textwidth} 
    \vspace{-10pt}
    \centering
    \includegraphics[width=\linewidth]{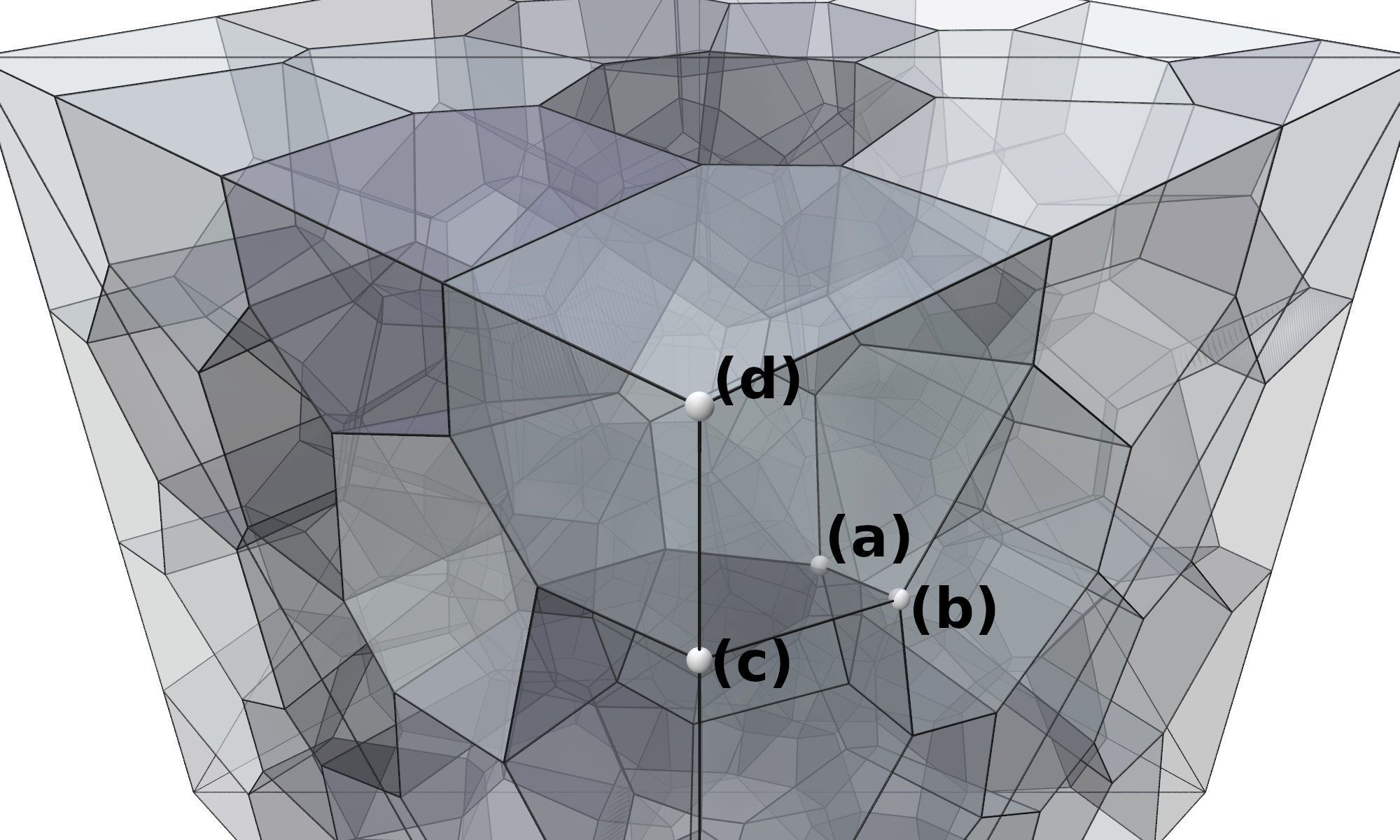}
    \label{fig:voronoi_vertices}
    \vspace{-20pt}
\end{wrapfigure}
(linearly independent) bisectors, (b) \textit{boundary face vertices}, \ie, intersections between two bisectors and one boundary facet, (c) \textit{boundary edge vertices}, \ie, intersections between one bisector and two boundary facets, and (d) \textit{simple boundary vertices}. The generating plane equations can be expressed as simple functions of the input degrees of freedom, \ie, the Voronoi sites and the vertices of the boundary surface. Given as the intersection of three planes, \ie, the solution of a $3\times3$ linear system, each 3D Voronoi vertex can be expressed in closed-form and differentiated with respect to the inputs.
\par
We derive as an example the closed-form expression for the unrestricted Voronoi vertex in 3D. Given four sites $\bx_i=(x_i,y_i,z_i)^T$, $i\in\{0,1,2,3\}$ of a Delaunay simplex, the plane equations of Voronoi bisectors are
\begin{equation}\label{eq:bisector}
    (x_j-x_i)x + (y_j - y_i)y + (z_j - z_i)z= \frac{1}{2}(x_j^2-x_i^2+y_j^2-y_i^2+z_j^2-z_i^2) \ .
\end{equation}
The intersection $\bx_c=(x_c,y_c,z_c)^T$ of three bisectors solves the linear system
\begin{equation*}\label{eq:voronoi_vertex_linear_system}
\begin{bmatrix} x_1 - x_0 & y_1 - y_0 & z_1 - z_0 \\x_2 - x_0 & y_2 - y_0 & z_2 - z_0 \\x_3 - x_0 & y_3 - y_0 & z_3 - z_0 \end{bmatrix} \begin{bmatrix} x_c \\ y_c \\ z_c \end{bmatrix} = \frac{1}{2} \begin{bmatrix} x_1^2-x_0^2+y_1^2-y_0^2+z_1^2-z_0^2 \\ x_2^2-x_0^2+y_2^2-y_0^2+z_2^2-z_0^2 \\ x_3^2-x_0^2+y_3^2-y_0^2+z_3^2-z_0^2 \end{bmatrix},
\end{equation*}
for which the solution can be written using the known closed-form expression for the $3\times3$ matrix inverse. This approach extends to all types of boundary vertices by substituting bisector plane equations in the linear system for plane equations of boundary facets. Furthermore, it is easily adapted for the power diagram by adding $\frac{1}{2}(w_i-w_j)$ to the right hand side of the bisector plane equation (\ref{eq:bisector}). In our implementation, code for evaluating these expressions and their derivatives is generated using a computer algebra system.
\subsection{Integration over Voronoi Cells}\label{sec:cell_integration}
\begin{figure}[H]
\centering
\begin{subfigure}{.24\linewidth}
  \centering
  \includegraphics[width=0.9\textwidth]{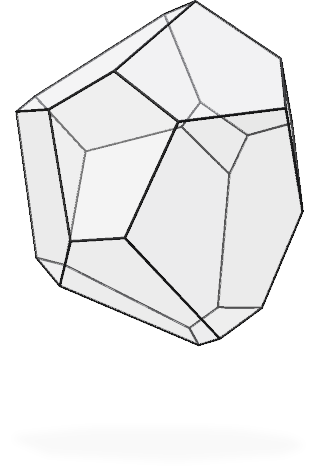}
\end{subfigure}
\begin{subfigure}{.24\linewidth}
  \centering
  \includegraphics[width=0.9\textwidth]{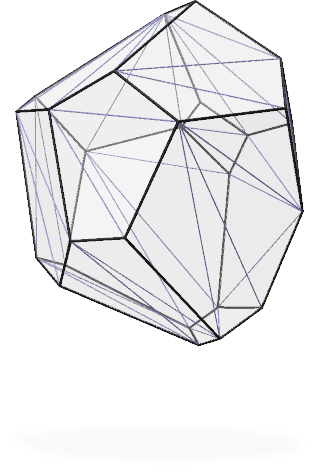}
\end{subfigure}
\begin{subfigure}{.24\linewidth}
  \centering
  \includegraphics[width=0.9\textwidth]{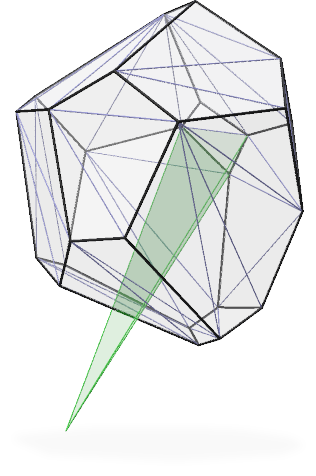}
\end{subfigure}
\begin{subfigure}{.24\linewidth}
  \centering
  \includegraphics[width=0.9\textwidth]{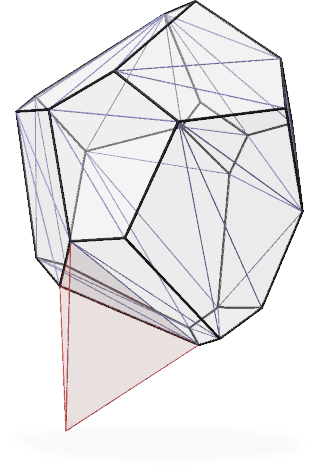}
\end{subfigure}
\caption{Volume integration over a 3D Voronoi cell. From left to right: (1) 3D Voronoi cell. (2) Face triangulation. (3) Integration tetrahedron formed by connecting a face triangle to the origin. Tetrahedra can have positive (3) or negative (4) volume.}
\label{fig:cell_integration}
\end{figure}

The mechanics of cell-based systems are driven by the energies of individual cells. Cell energies relate to properties such as surface area and volume, which are computed by integrating over the cell geometry. Volume integrals are also used to compute cell centroids and moments of inertia.
Integration over a polyhedral Voronoi cell is performed piecewise by dividing the surface and volume into simplices. A surface integral requires triangulating each facet of the cell and then summing the per-triangle values. A tetrahedralization of the volume is, however, unnecessary for computing volume integrals. Rather, tetrahedra are constructed by joining each surface triangle to the origin as shown in Fig. \ref{fig:cell_integration}. 
In addition to avoiding an explicit tetrahedralization, including the origin in each tetrahredron simplifies the simplex integral to a function of three vertices instead of four. Importantly, this reduces the number of first and second derivatives required per simplex. A generic approach for volume integration in these tetrahedra is given in App.~\ref{sec:tetrahedron_integration}.
\subsection{Energy Formulation}

We simulate cell mechanics in the implicit Voronoi model by considering a potential energy of the form
\begin{equation}\label{eq:energy_form_no_boundary}
    E = F(\bc, \bx(\bc)),
\end{equation}
where $\bc$ is a vector containing the degrees of freedom of all Voronoi sites, and $\bx$ is a vector of spatial coordinates of all Voronoi vertices, defined implicitly in terms of $\bc$. In general, an energy depending only on $\bx$ is sufficient to describe physical interactions between cells, and explicit dependence on $\bc$ is used for regularization.
The generic gradient of this energy is \begin{equation}\label{eq:energy_grad_no_boundary}
    \frac{\dd E}{\dd \bc} = \frac{\partial F}{\partial \bx}\frac{\dd \bx}{\dd \bc} + \frac{\partial F}{\partial \bc} \ ,
\end{equation}
and the Hessian follows as
\begin{equation}\label{eq:energy_hess_no_boundary}
\begin{aligned}
\frac{\dd^2 E}{\dd \bc^2}
& = \left(\frac{\dd \bx}{\dd \bc}\right)^\top \frac{\partial^2 F}{\partial \bx^2} \frac{\dd \bx}{\dd \bc} + \left(\frac{\dd \bx}{\dd \bc}\right)^\top \frac{\partial^2 F}{\partial \bx \partial \bc} \\
& + \frac{\partial^2 F}{\partial \bc \partial \bx} \frac{\dd \bx}{\dd \bc} + \sum_i \left(\frac{\partial F}{\partial \bx_i}\frac{\dd^2 \bx_i}{\dd \bc^2}\right) + \frac{\partial^2 F}{\partial \bc^2} \ .
\end{aligned}
\end{equation}
This flexible formulation permits any differentiable energy function and is applicable to a wide range of cell-based systems. The same computational method can be used to simulate foam as well as biological tissue, as demonstrated in Sec.~\ref{sec:results}.

\subsection{Boundary Coupling}\label{sec:boundary_coupling}

Many cellular systems interface with external bodies or free space. Biological tissues grow freely or within deformable membranes, while engineering designs that utilize cellular structures must support external loads.
To handle these external interactions, we define the bounding domain of the Voronoi diagram by a triangular mesh with vertices $\bv(\bp)$, where $\bp$ are boundary degrees of freedom. We let $\by = (\bc, \bp)$ denote the vector holding all degrees of freedom of the coupled system and consider a total potential energy
\begin{equation}\label{eq:energy_form_with_boundary}
    E = F\left(\bc, \bx\left(\bc, \bv(\bp)\right)\right) + F_B(\bp) \ ,
\end{equation}
where $F$ is the potential energy of the cells and $F_B$ is an additional boundary energy which may encode, \eg, the elastic potential of an enclosing membrane. As for the static boundary formulation, we compute the first and second derivatives of the above expression to use in gradient-based optimization. The gradient with respect to $\bp$ is given by
\begin{equation}\label{eq:energy_grad_with_boundary}
\frac{\dd E}{\dd \bp} = \frac{\partial F}{\partial \bx}\frac{\partial \bx}{\partial \bv}\frac{\dd \bv}{\dd \bp} + \frac{\dd F_B}{\dd \bp} \ .
\end{equation}
The additional blocks of the Hessian matrix are given in App. \ref{sec:appHessianBoundary}. 
\begin{figure*}
\centering
\begin{subfigure}{.99\linewidth}
  \centering
  \includegraphics[width=0.99\textwidth, trim={0 8.7cm 0 9.8cm}, clip]{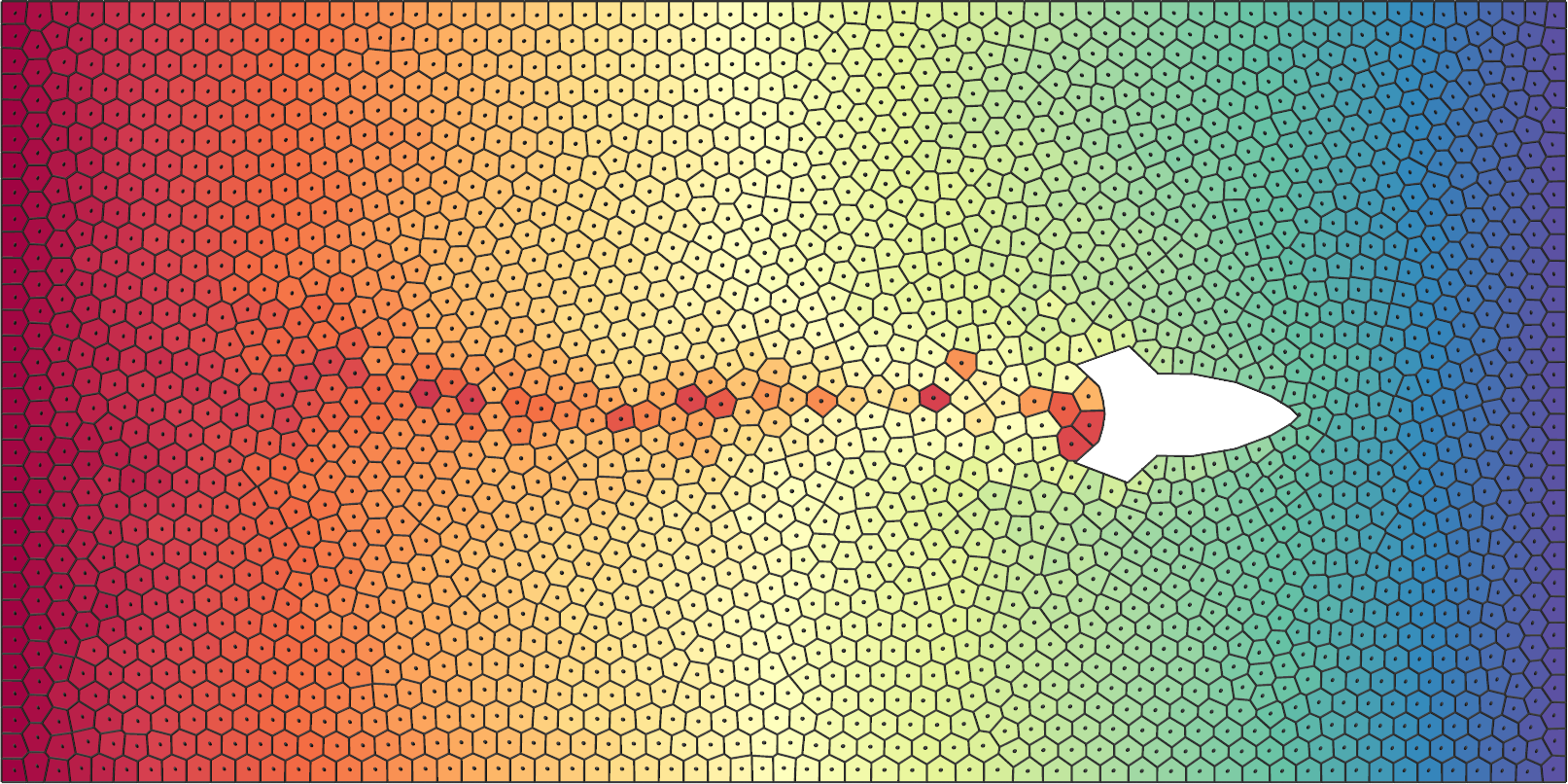}
\end{subfigure}
\begin{subfigure}{.99\linewidth}
  \centering
  \includegraphics[width=0.99\textwidth, trim={0 12.9cm 0 5.6cm}, clip]{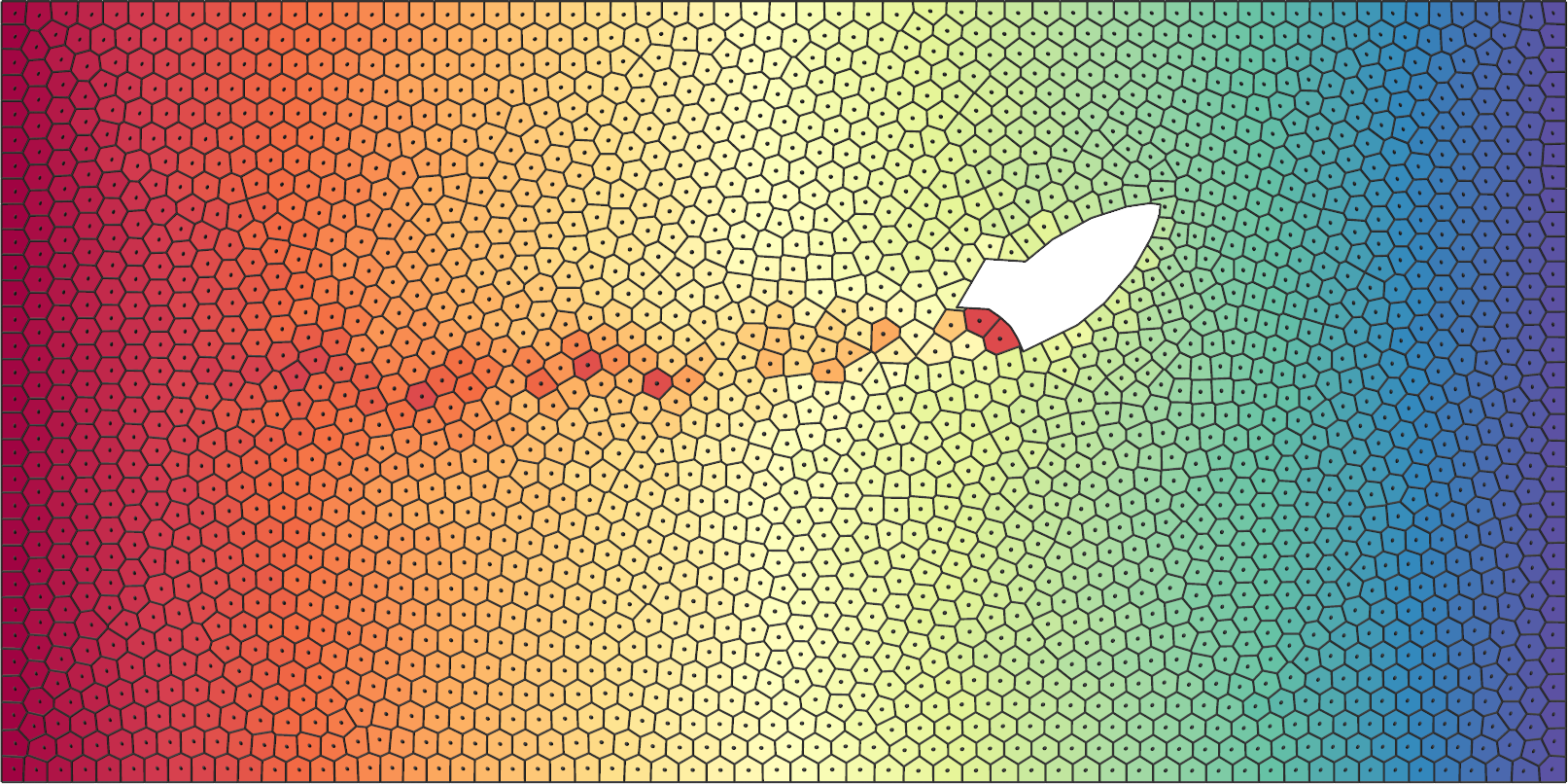}
\end{subfigure}
\caption{Two simulations of a rigid body propelled through a tissue-like assembly of elastic Voronoi cells. Cells are colored based on starting position. Turbulence-like effects result in irregular displacement of cells from their starting positions. The asymmetric shape of the second body results in a curved trajectory due to forces from neighboring cells.}
\label{fig:rocket_ship}
\end{figure*}
%
This boundary coupling formulation lends extreme flexibility to the model, enabling systems involving elastic membranes, free surfaces, and rigid bodies in addition to fixed boundaries.
\subsection{Smoothness}
In the general configuration, where each Voronoi vertex in $m$ dimensions is of degree $m+1$, the topology of the Voronoi diagram is constant within a neighborhood of the given state, and the vertex coordinates $\bx$ are a locally smooth function of the Voronoi sites $\bc$. 
States with higher-degree vertices, which occur during topological transitions, have undefined $\frac{\dd \bx}{\dd \by}$ which gives rise to non-smooth energies. Cell volume (area in 2D) and other volume integrals of smooth functions are $C^1$-continuous across these transitional states, though surface area (perimeter in 2D) is $C^0$ as noted in~\cite{Bogosel2022} for the 2D case.
\par
If cell energy is monotonic in surface area and the domain boundary is convex, the gradient discontinuities are strictly concave, forming local maxima in the energy landscape which do not hinder minimization. Otherwise, the model may fail to converge using such energy functions. Our simulations with non-convex domain boundaries (Secs.~\ref{sec:rocket_ship} and ~\ref{sec:cleavage}) use $C^1$-continuous energies and therefore avoid these problems.
\par
Degenerate configurations exist with lower-order discontinuities, \eg when a parallel boundary facet and Voronoi bisector overlap (discussed in~\cite{CVT2009}) or when multiple Voronoi sites coincide. However, we have not observed these states during simulation and optimization in any of our examples.
\subsection{Dynamics}\label{sec:dynamics}
We have so far considered systems at static equilibrium in which the net internal force vanishes,
\begin{equation}\label{eq:force_energy_gradient}
F(\by) = -\frac{\dd E}{\dd \by}=\mathbf{0}\ .
\end{equation}
We now relax the equilibrium constraint to consider dynamic cell-based systems governed by Newton's second law,
\begin{equation}\label{eq:newton_second_law}
F(\by) = \mathbf{M}\frac{\dd^2 \by}{\dd t^2} \ ,
\end{equation}
where $\mathbf{M}$ is a (typically diagonal) mass matrix. We additionally introduce viscous forces $F_{f}(\by) = -\boldsymbol{\eta} \frac{\dd \by}{\dd t}$, where $\boldsymbol{\eta}$ is a diagonal viscosity matrix.
The resulting equations of motion are
\begin{equation}\label{eq:eq_of_motion_general}
\mathbf{M}\ddot{\by} + \boldsymbol{\eta}\dot{\by} + \frac{\dd E}{\dd \by} = \bZero \ ,
\end{equation}
where $\dot{\by}$ and $\ddot{\by}$ denote first and second derivatives of $\by$ with respect to time. Unless otherwise noted, we approximate these derivatives using standard finite-differencing, \ie, 
\begin{equation}\label{eq:eq_of_motion_first_order}
\begin{gathered}
\dot{\by} \approx \frac{1}{h}(\by_{k+1}-\by_{k}) \ , \\
\ddot{\by} \approx \frac{1}{h^2}(\by_{k+1}-2\by_{k}+\by_{k-1}) \ . \\
\end{gathered}
\end{equation}
If the energy function is at least $C^1$-continuous, quadratic convergence can be achieved using a second-order discretization (BDF2) for the accelerations,
\begin{equation}\label{eq:eq_of_motion_second_order}
\ddot{\by} \approx \frac{1}{2h^2}(3\by_{k+1}-7\by_{k}+5\by_{k-1}-\by_{k-2}) \ .
\end{equation}
Evidence of quadratic convergence under refinement of the simulation time step is shown in App.~\ref{sec:convergence_analysis}.
The equations of motion can be solved for the next state $\by_{k+1}$ via the optimization problem
\begin{equation}\label{eq:dynamics_objective}
\by_{k+1} = \argmin_{\by} \frac{a_2}{2} \ddot{\by}^\top \mathbf{M} \ddot{\by} + \frac{a_1}{2} \dot{\by}^\top \boldsymbol{\eta} \dot{\by} + E(\by) \ ,
\end{equation}
where $a_1,a_2$ are the $\by_{k+1}$-coefficients of the first and second finite-difference derivatives. Many cell-based systems operate at low Reynolds numbers, where inertia is dominated by viscous effects. For such cases, the momentum term can be neglected, leaving
\begin{equation}\label{eq:dynamics_objective_no_momentum}
\by_{k+1} = \argmin_{\by} \frac{a_1}{2} \dot{\by}^\top \boldsymbol{\eta} \dot{\by} + E(\by)\ .
\end{equation}
\begin{figure*}
\centering
\begin{subfigure}{.32\linewidth}
  \centering
  \includegraphics[width=0.95\textwidth]{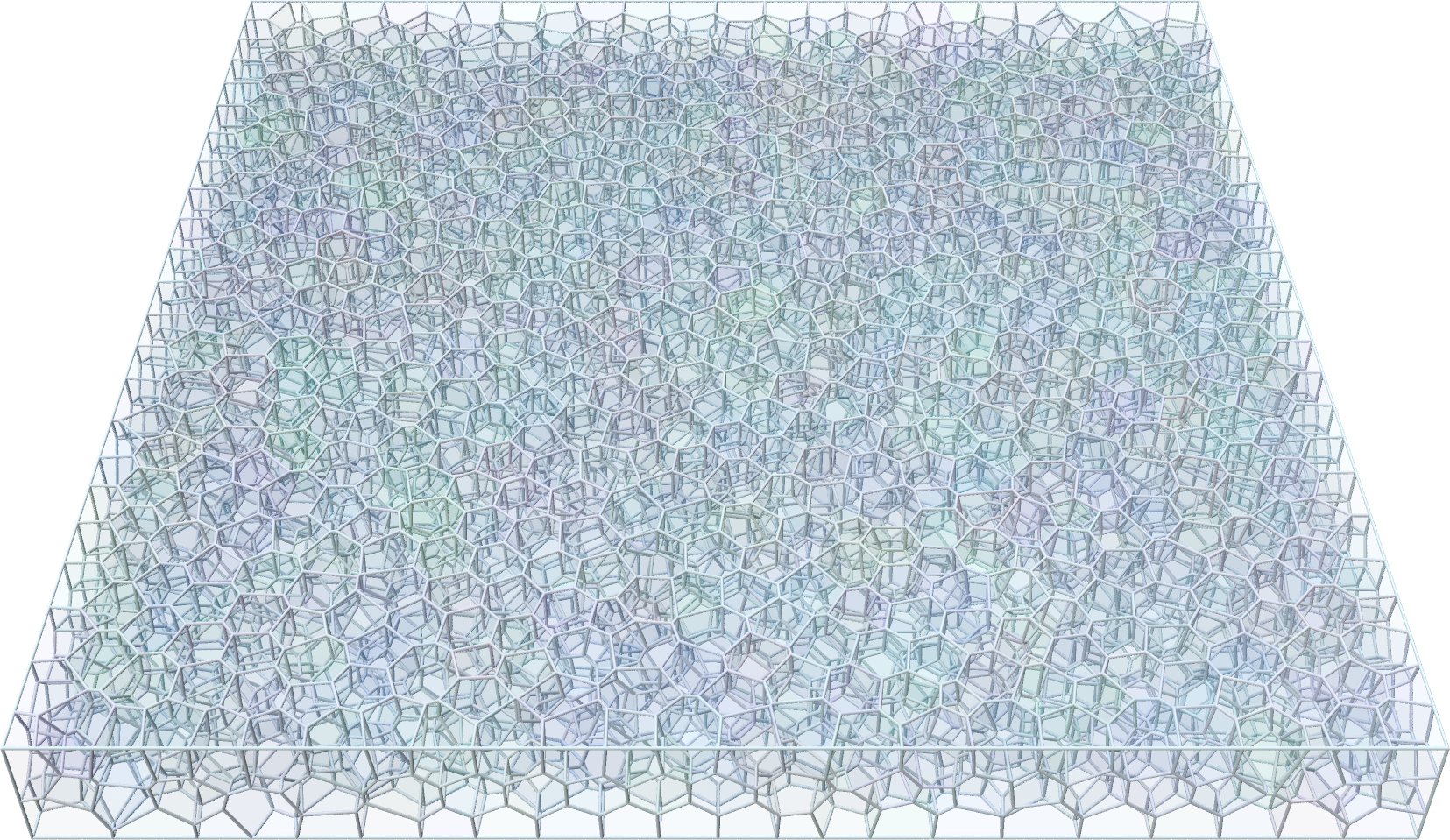}
  \caption{Frame 0, $n=2000$.}
\end{subfigure}
\begin{subfigure}{.32\linewidth}
  \centering
  \includegraphics[width=0.95\textwidth]{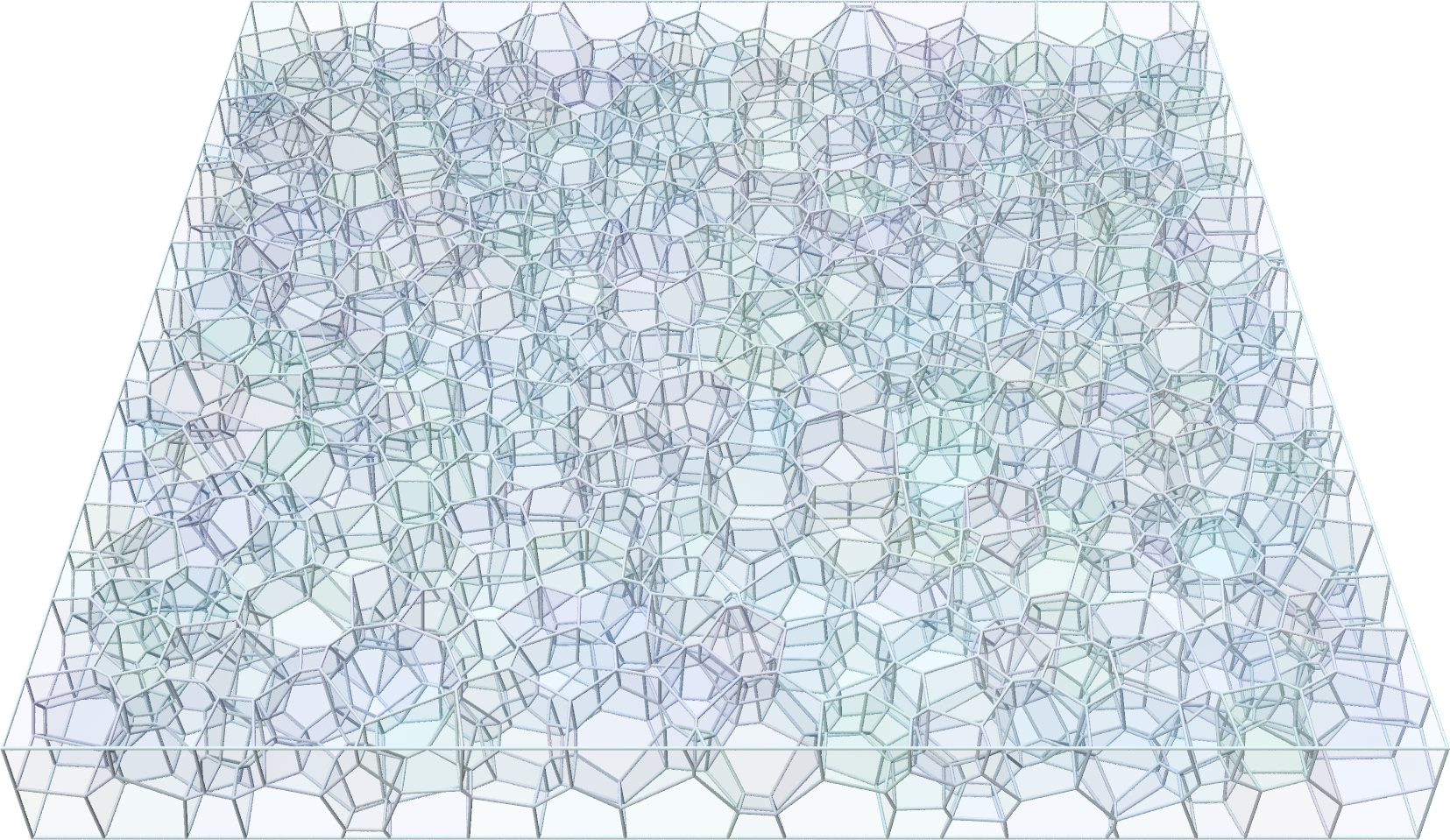}
  \caption{Frame 31, $n=794$.}
\end{subfigure}
\begin{subfigure}{.32\linewidth}
  \centering
  \includegraphics[width=0.95\textwidth]{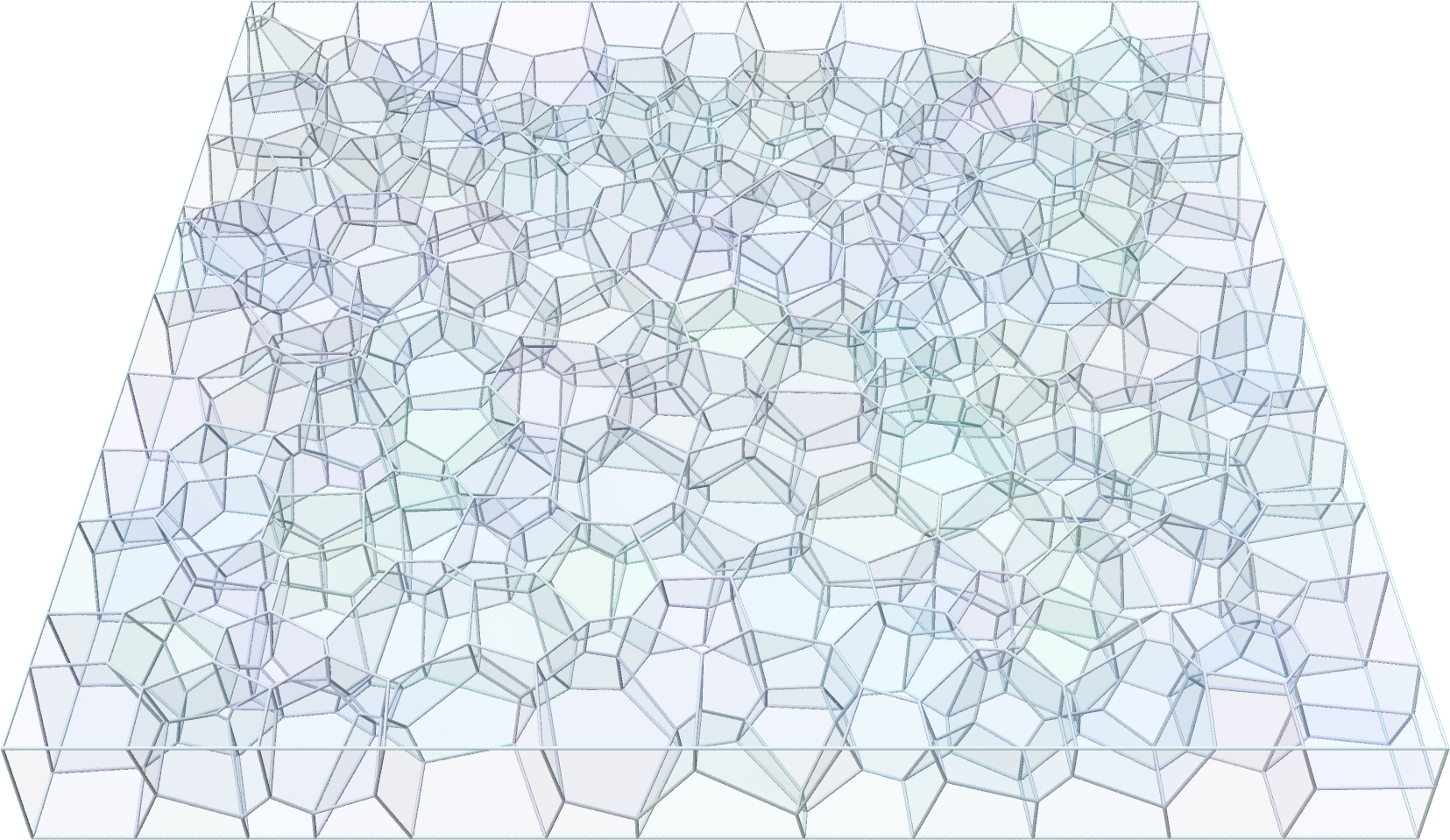}
  \caption{Frame 50, $n=282$.}
\end{subfigure}
\begin{subfigure}{.32\linewidth}
  \centering
  \includegraphics[width=0.95\textwidth]{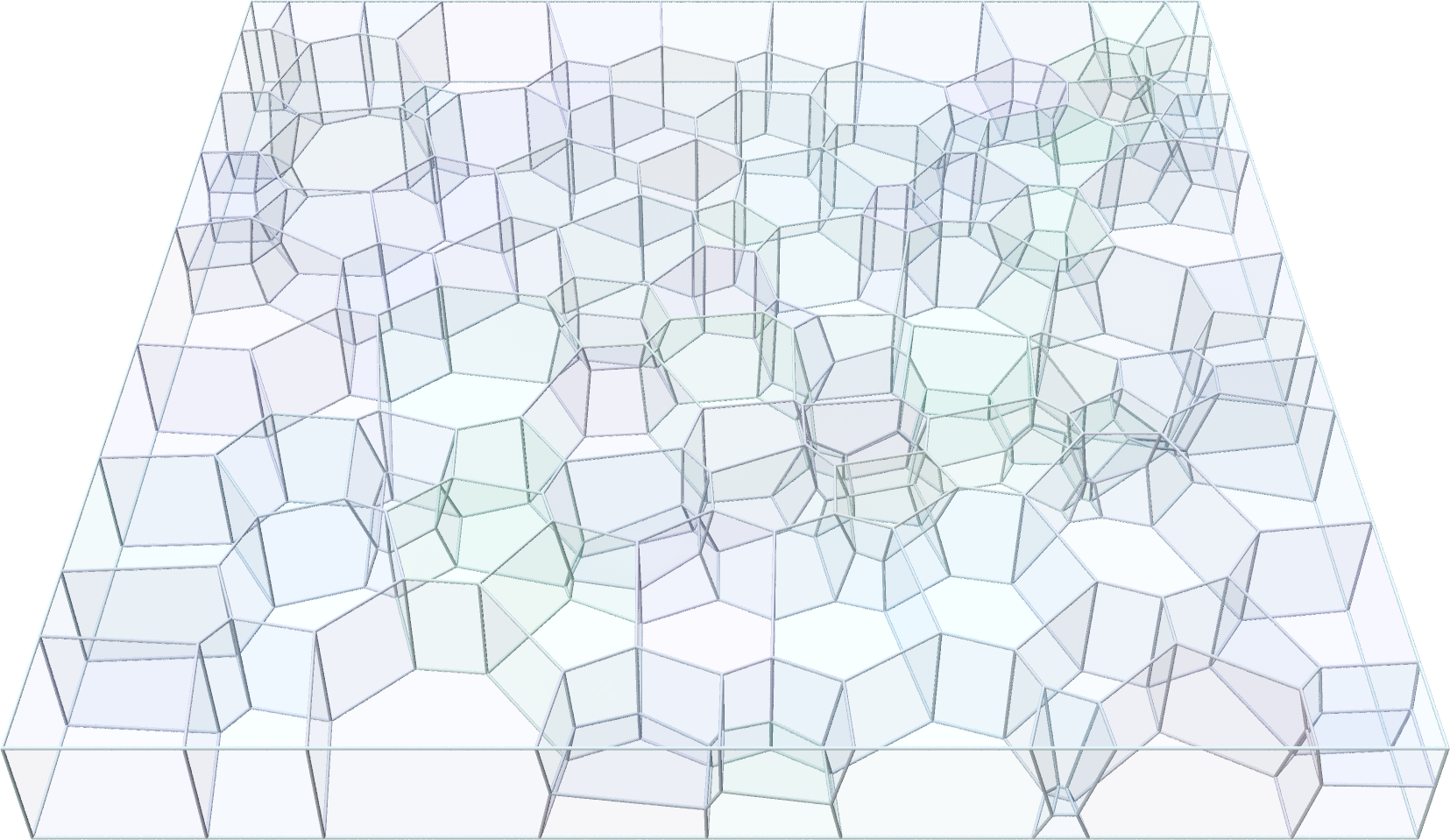}
  \caption{Frame 65, $n=121$.}
\end{subfigure}
\begin{subfigure}{.32\linewidth}
  \centering
  \includegraphics[width=0.95\textwidth]{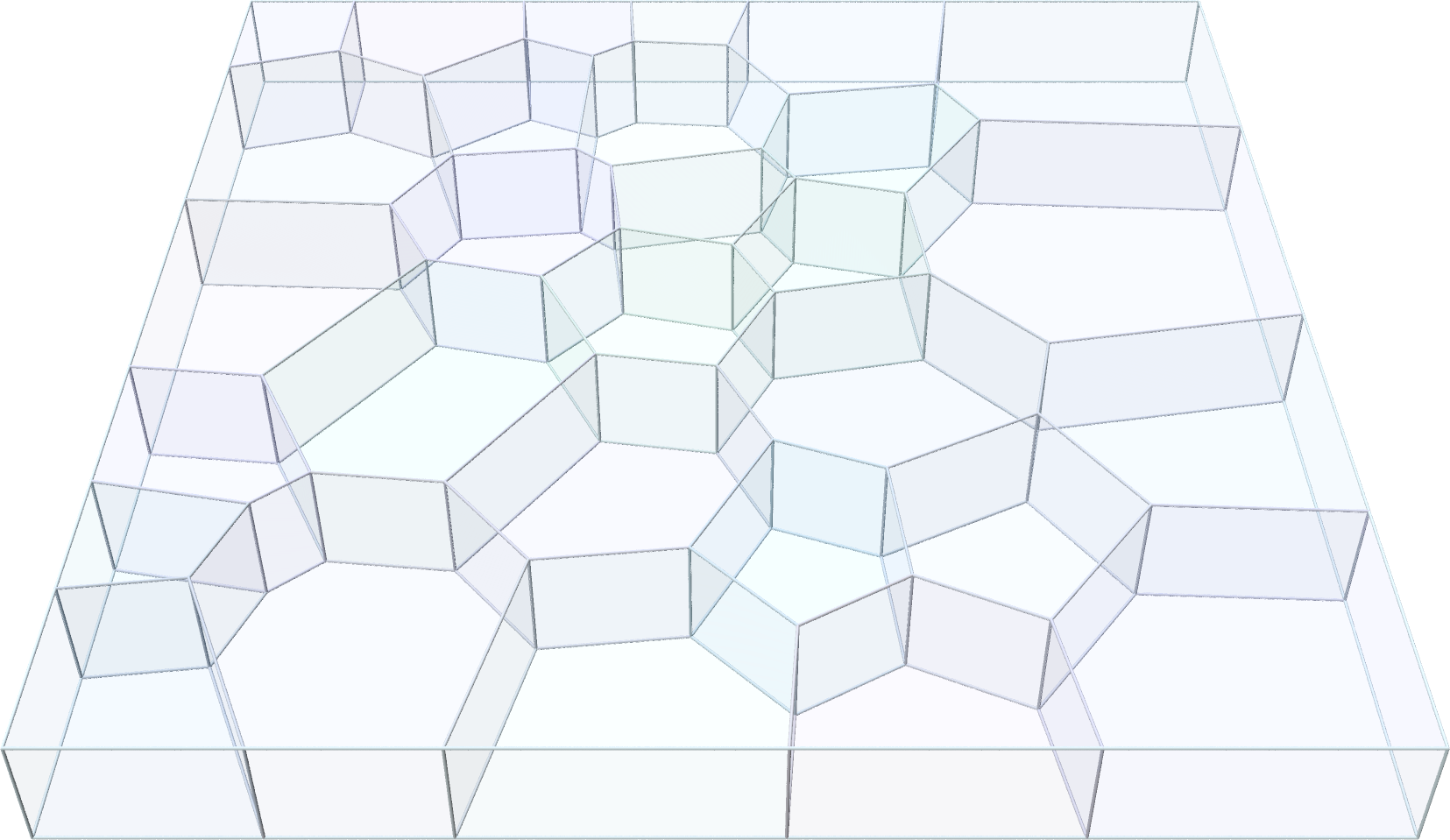}
  \caption{Frame 110, $n=29$.}
\end{subfigure}
\begin{subfigure}{.32\linewidth}
  \centering
  \includegraphics[width=0.95\textwidth]{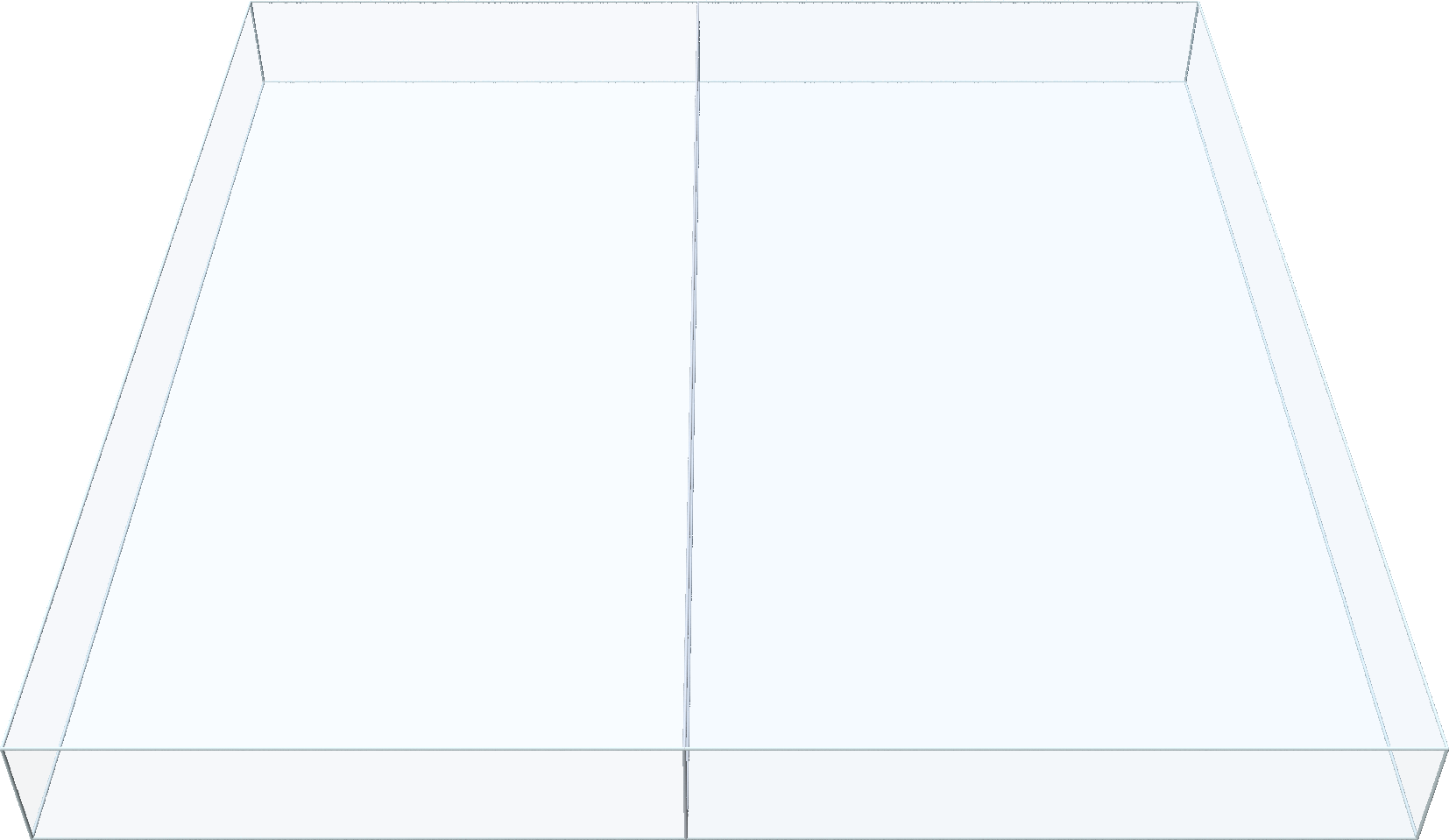}
  \caption{Final equilibrium, $n=2$.}
\end{subfigure}
\caption{Coarsening of an initially monodisperse dry foam in a flattened box. 2000 cells collapse to a two-cell equilibrium in 350 frames.}
\label{fig:coarsening}
\end{figure*}
\subsection{Equilibrium-Constrained Optimization}\label{sec:eq_constrained_optimization}
In addition to forward simulation, our approach can be extended to solve inverse simulation tasks formulated as
equilibrium-constrained optimization problems. The goal is then to find optimal parameter values for a cell-based system such that the corresponding equilibrium state best satisfies a given objective. 
We demonstrate in Sec.~\ref{sec:results_image_matching} that this framework is useful for characterizing real-world cell-based systems from images. 
\par
A general equilibrium-constrained optimization problem reads
\begin{equation}\label{eq:eq_constrained_problem}
\min_{\by, \bu} L(\by, \bu) \quad \text{s.t.} \quad \frac{\partial}{\partial \by}E(\by, \bu) = \bZero \ ,
\end{equation}
where $L$ is an objective function, $\by$ are dynamic degrees of freedom, and $\bu$ are optimization variables. The system is constrained to be at equilibrium, hence the state $\by$ is a function of $\bu$. The gradient is given by
\begin{equation}\label{eq:eq_constrained_objective_gradient}
\frac{\dd L}{\dd \bu} = \frac{\partial L}{\partial \by}\frac{\dd \by}{\dd \bu} + \frac{\partial L}{\partial \bu} \ ,
\end{equation}
where the simulation derivatives $\frac{\dd \by}{\dd \bu}$ can be computed using the implicit function theorem,
\begin{equation}
\label{eq:eq_constrained_sensitivity_analysis}
\frac{\dd}{\dd \bu}\frac{\partial}{\partial \by}E(\by, \bu)  = \bZero 
\ \rightarrow \ 
\frac{\dd \by}{\dd \bu} = -\left(\frac{\partial^2 E}{\partial \by^2}\right)^{-1} \frac{\partial^2 E}{\partial \by \partial \bu} \ .
\end{equation}
Having evaluated the objective gradient with respect to the optimization variables $\bu$, we can minimize the $L$ using first-order or quasi-Newton methods. Whenever the optimization parameters are updated during this process, we recompute the equilibrium state $\by(\bu)$ and ensure monotonic decrease of the objective using a backtracking line search. 
\subsection{Implementation}
The method is implemented in C\texttt{++} using the Eigen library~\cite{eigenweb} for matrix and vector operations, and the CHOLMOD solver~\cite{cholmod} for linear systems. Geometry processing libraries CGAL~\cite{cgal:y-t2-23a} and Geogram~\cite{geogram} are used in Voronoi diagram generation. Finally, we use Polyscope~\cite{polyscope} for visualization and rendering of figures.
Our code can be found at \href{https://github.com/lnumerow-ethz/VoronoiCellSim}{https://github.com/lnumerow-ethz/VoronoiCellSim}.

\section{Results}\label{sec:results}
\subsection{Intercellular Navigation with Rigid Body}\label{sec:rocket_ship}
In this first 2D experiment, we simulate the propulsion of a rigid body through a tissue-like assembly of elastic Voronoi cells. The cell energy is given by
\begin{equation}\label{eq:rocket_ship_energy}
    E = a_0(A-\bar{A})^2 + a_1\int_{x\in\mathcal{R}}\|x-\bar{x}\|^2 dA + a_2\|c-\bar{x}\|^2 \ ,
\end{equation}
where $A$ and $\bar{A}$ are the current and target area of the cell, $\bar{x}$ is its centroid, and $c$ is the Voronoi site position. In this and the following examples, the final term is a regularizer on the distance between the site and its cell centroid. The rigid body, modelled as an internal clipping geometry with translation and rotation degrees of freedom (see Sec.~\ref{sec:boundary_coupling}), is subject to a constant force $F_x$ to the right.
We use a dynamics model with momentum (Eq.~\ref{eq:dynamics_objective}) to resolve turbulence-like effects in the system, the results of which can be observed in Fig.~\ref{fig:rocket_ship}.
The curved trajectory of the rigid body in the second example is a result of its asymmetric shape and forces applied by the surrounding cells. This demonstrates the dynamic coupling between the cells and boundary degrees of freedom.

For this and the following experiments, refer to App.~\ref{sec:experiment_details} for parameter values and additional details for reproducibility.
\subsection{Foam Coarsening}\label{sec:coarsening}
Due to pressure differences between bubbles in an equilibriated liquid foam, gas diffuses slowly through the interfaces from smaller, higher-pressure bubbles into larger, lower-pressure ones. As a result, smaller foam cells collapse and the total number of foam cells in the system decreases over time in a process known as \textit{coarsening}. 
This phenomenon has seen extensive research~\cite{VedelLarsen2010, Thomas2015} due to its detrimental effect in applications such as enhanced oil recovery.
Foam coarsening presents a challenge for vertex models due to the topological changes which result from the collapse and removal of a cell, particularly in 3D. Existing 2D vertex models \cite{VedelLarsen2010} rely on heuristics to remove collapsing cells and reconstruct topology, resulting in discontinuous changes to neighboring cell volumes and to the system energy.
\par
We simulate dry foam coarsening using a momentumless dynamic model as described in Eq. \ref{eq:dynamics_objective_no_momentum}. Each site has five degrees of freedom: three spatial coordinates $(x,y,z)$, a power diagram weight $w$ and a volume target $\bar{V}$ representing the mass of air contained in the cell. The energy of a cell is
\begin{equation}\label{eq:coarsening_energy}
    E = a_0\left(\frac{V}{\bar{V}}-1\right)^2 + a_1A + a_2\|c-\bar{x}\|^2 \ ,
\end{equation}
where $V$ and $A$ are the volume and surface area, $\bar{x}$ is the centroid and $c$ is the Voronoi site position. The viscosity coefficients (diagonal entries of $\boldsymbol{\eta}$ in Eq. (\ref{eq:dynamics_objective_no_momentum})) of the site degrees of freedom are very small, such that the system evolves quasi-statically. The larger viscosity coefficient, corresponding to the volume target $\bar{V}$, represents the resistance to diffusion of air through the foam interfaces.
Fig.~\ref{fig:coarsening} shows the result of a large-scale 3D coarsening simulation beginning with $n=2000$ monodisperse foam cells. Variation of power diagram weights in our differentiable Voronoi model allows cells to collapse smoothly to zero volume. Furthermore, our model implicitly performs the complete topological restructuring of the local neighborhood that results from each cell collapse. It should be noted that the geometric accuracy of this simulation, and other foam simulations involving wide variation of cell sizes and pressures, is limited due to the lack of curved interfaces.
\subsection{Biological Tissue Growth}\label{sec:cleavage}
Embryonic \textit{cleavage} is the process in which a complex organism grows from a single cell via repeated cell division. It involves an exponential increase in the number of cells and therefore a rapid increase in the complexity of an organism. Cell \textit{proliferation}, a similar process that involves the growth and subsequent division of cells, is another primary driver of tissue development. Simulation of such processes is a common use case for computational models~\cite{PalaCell2021, PhysiCell3D}.
Our first experiment simulates embryonic cleavage in a spherical membrane, beginning with a single Voronoi cell. By the dynamic coupling formulation described in Sec.~\ref{sec:boundary_coupling}, the domain boundary is deformable and subject to its own elastic forces, with a boundary mesh edge of length $\ell$ having energy $E_\text{B}=k\ell^2$. The cell energy is
\begin{equation}\label{eq:cleavage_energy}
    E = a_0\left(\frac{V}{\bar{V}}-1\right)^2 + \frac{a_1}{\bar{V}^{4/3}}\int_{x\in\mathcal{R}}\|x-\bar{x}\|^2 dV + \frac{a_2}{\bar{V}^{2/3}}\|c-\bar{x}\|^2,
\end{equation}
where $V$ and $\bar{V}$ are the current and rest volumes, $\bar{x}$ is the centroid and $c$ is the Voronoi site position. We simulate the evolution of the system using a momentumless dynamic model as described in Eq. (\ref{eq:dynamics_objective_no_momentum}). After every $k$ simulation time steps, each Voronoi site is replaced by two daughter sites at positions $c \pm \beta \bar{V}^{1/3} \bn$, where $\bn$ is the unit normal vector to the cleavage plane. The first three cleavages are orthogonal, after which the cleavage planes are chosen randomly. The daughter sites are initialized with the same power diagram weight as the parent site, and rest volume $\frac{1}{2}\bar{V}_{\text{parent}}$.
The system is simulated over 700 time steps with a division every $T=60$ frames, resulting in a final cell count of 4096. In Fig.~\ref{fig:cleavage_membrane}, the simulated embryo is compared to frames from~\cite{vanIjken} depicting embryonic development of a salamander. While this comparison is not intended to be quantitative, it shows that our method is able to produce simulations qualitatively similar to real-world cellular systems undergoing many topological changes.

\begin{figure}[h]
\centering
\begin{subfigure}{.32\linewidth}
  \centering
  \includegraphics[width=0.9\textwidth]{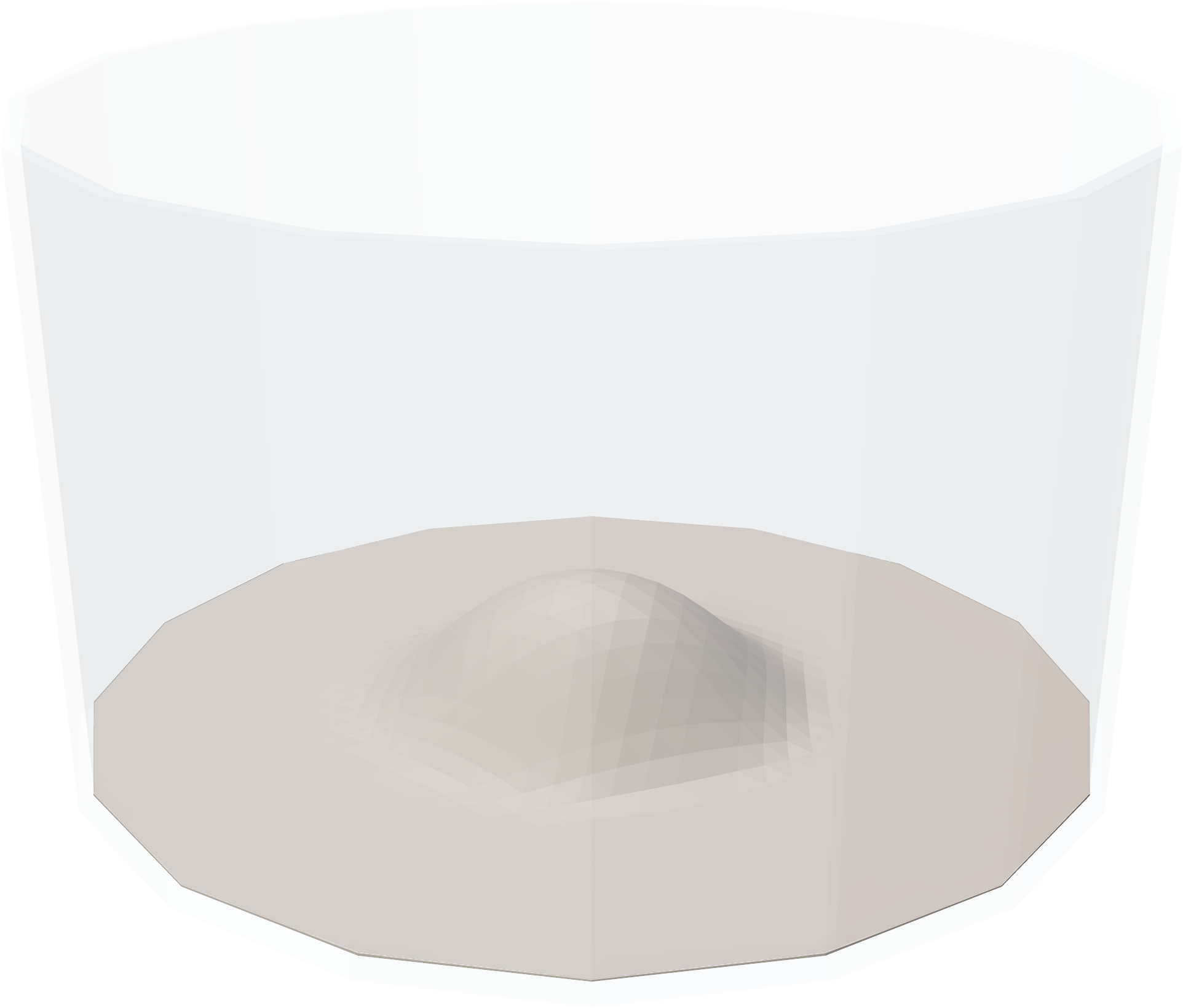}
  \caption{Frame 0, $n=1$.}
\end{subfigure}
\begin{subfigure}{.32\linewidth}
  \centering
  \includegraphics[width=0.9\textwidth]{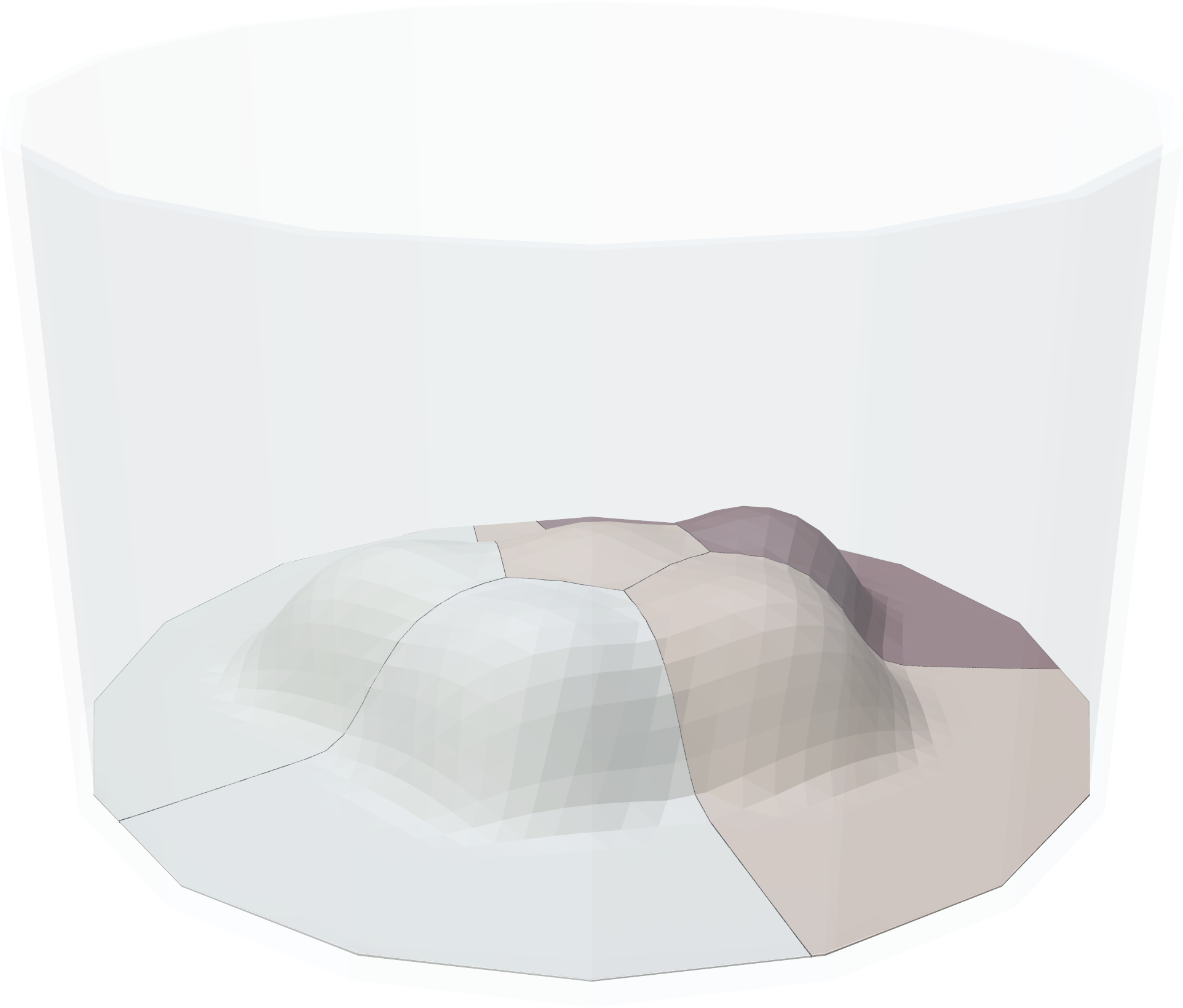}
  \caption{Frame 100, $n=5$.}
\end{subfigure}
\begin{subfigure}{.32\linewidth}
  \centering
  \includegraphics[width=0.9\textwidth]{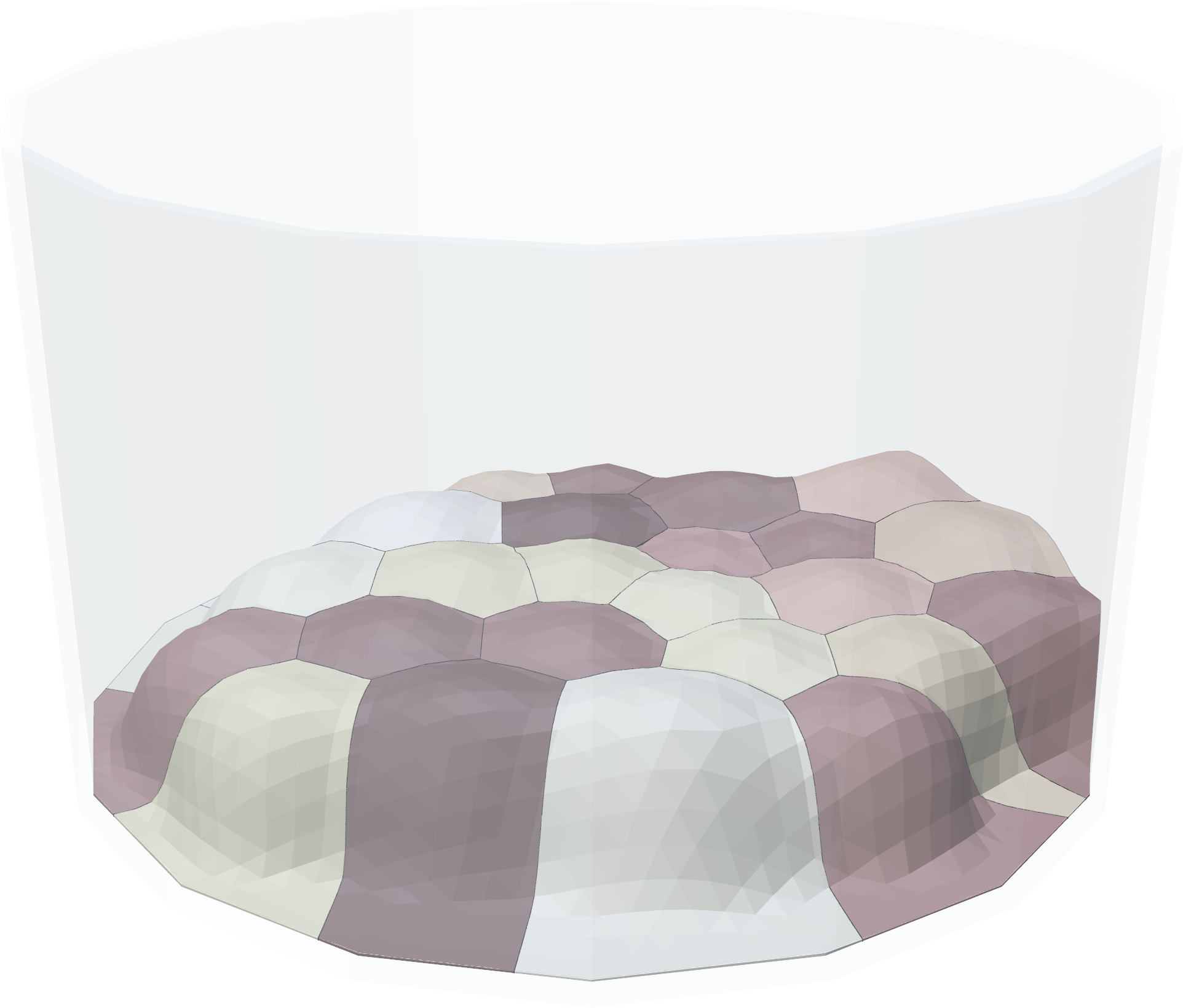}
  \caption{Frame 250, $n=24$.}
\end{subfigure}
\begin{subfigure}{.32\linewidth}
  \centering
  \includegraphics[width=0.9\textwidth]{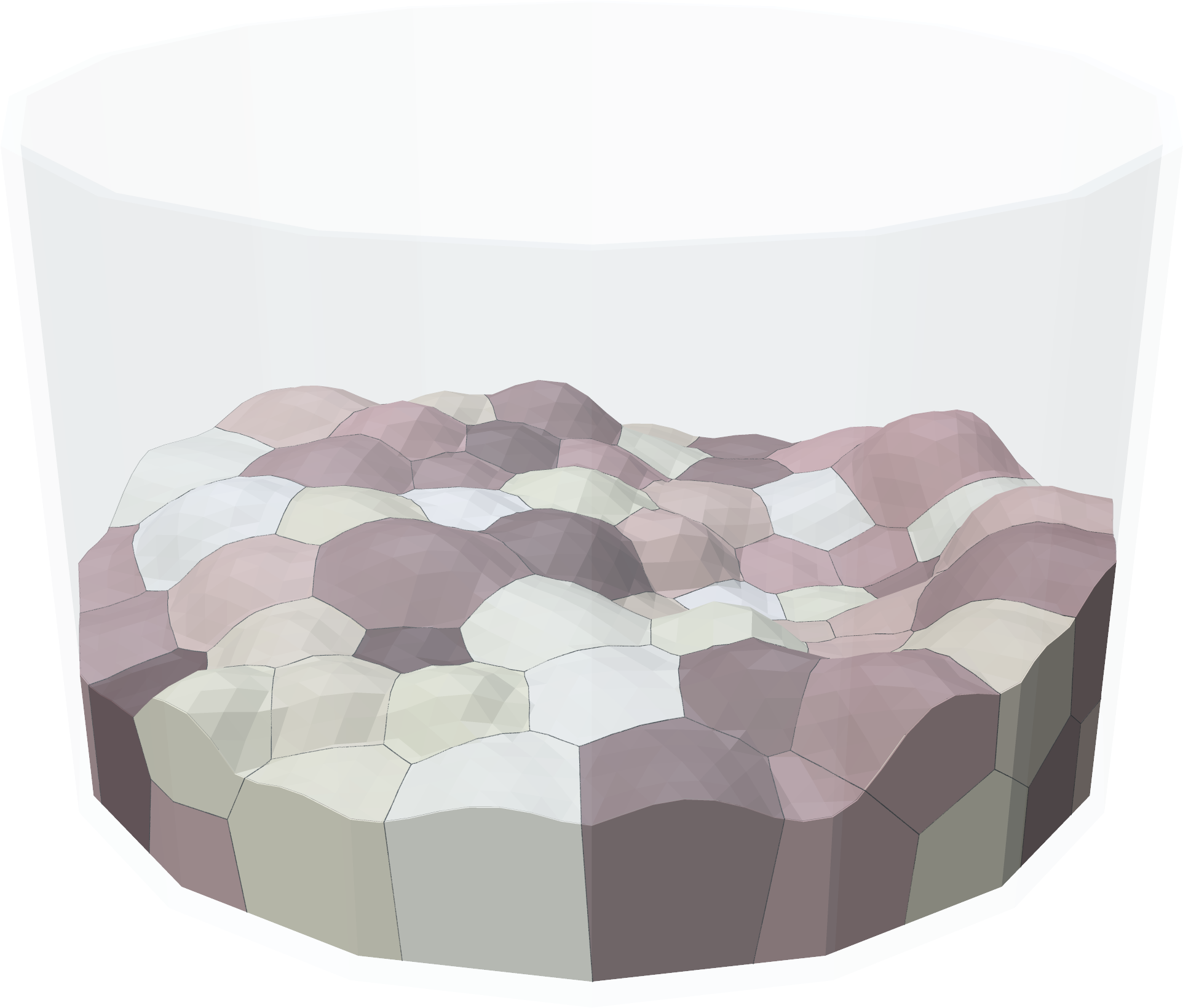}
  \caption{Frame 450, $n=87$.}
\end{subfigure}
\begin{subfigure}{.32\linewidth}
  \centering
  \includegraphics[width=0.9\textwidth]{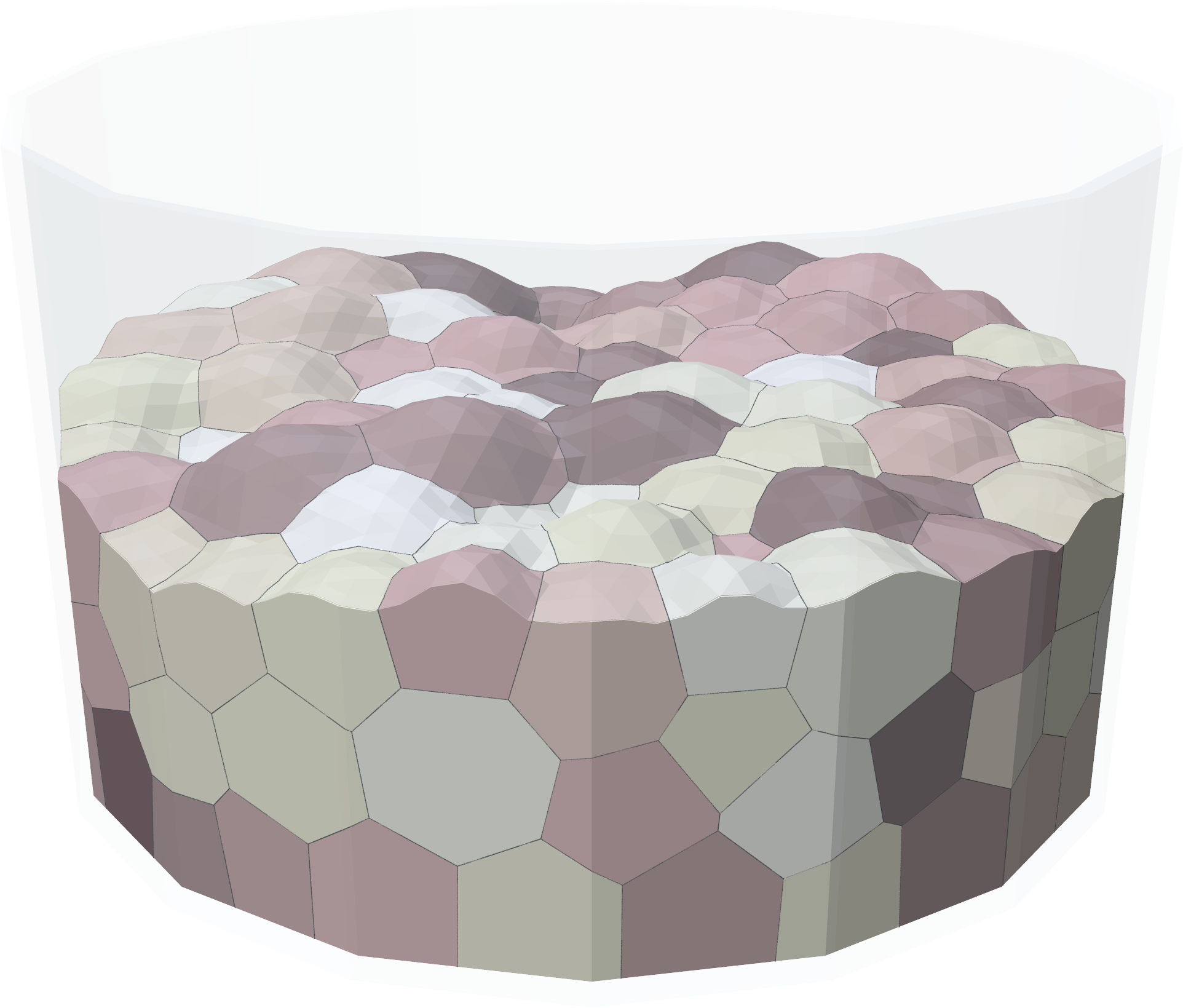}
  \caption{Frame 736, $n=250$.}
\end{subfigure}
\begin{subfigure}{.32\linewidth}
  \centering
  \includegraphics[width=0.9\textwidth]{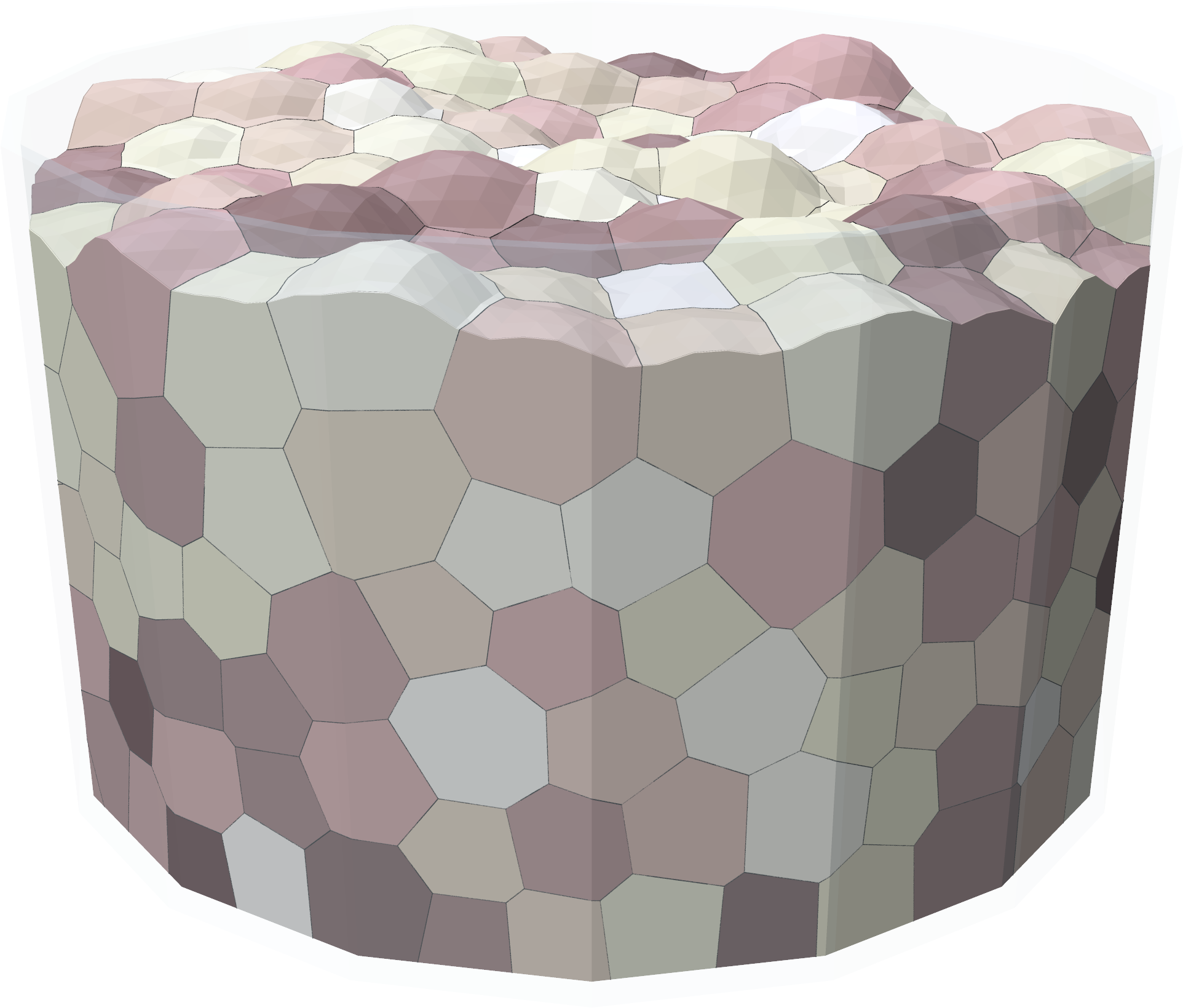}
  \caption{Frame 1100, $n=620$.}
\end{subfigure}
\caption{Cell proliferation in a cylindrical container. The free boundary is coupled to the cells using a weak elastic membrane model. Random cell divisions induce frequent and irregular topology changes throughout the simulation.}
\label{fig:cleavage_cylinder}
\end{figure}

A second experiment simulates cell proliferation in a cylindrical container of radius 1, with an added gravity term and free upper surface (Fig.~\ref{fig:cleavage_cylinder}). After each simulation step of size $h$, each cell divides with probability $p = \alpha\bar{V}h$, where $\alpha$ is the proliferation rate of the tissue. The daughter sites are initialized with rest volume $\frac{1}{2}\bar{V}_{\text{parent}}$ increasing linearly to $\gamma\bar{V}_{\text{parent}}$ over time period $\tau$.

%
\subsection{Characterization of Foam from Image}\label{sec:results_image_matching}
\begin{figure}[h]
\centering
\begin{subfigure}{.32\linewidth}
  \centering
  \includegraphics[width=0.99\textwidth]{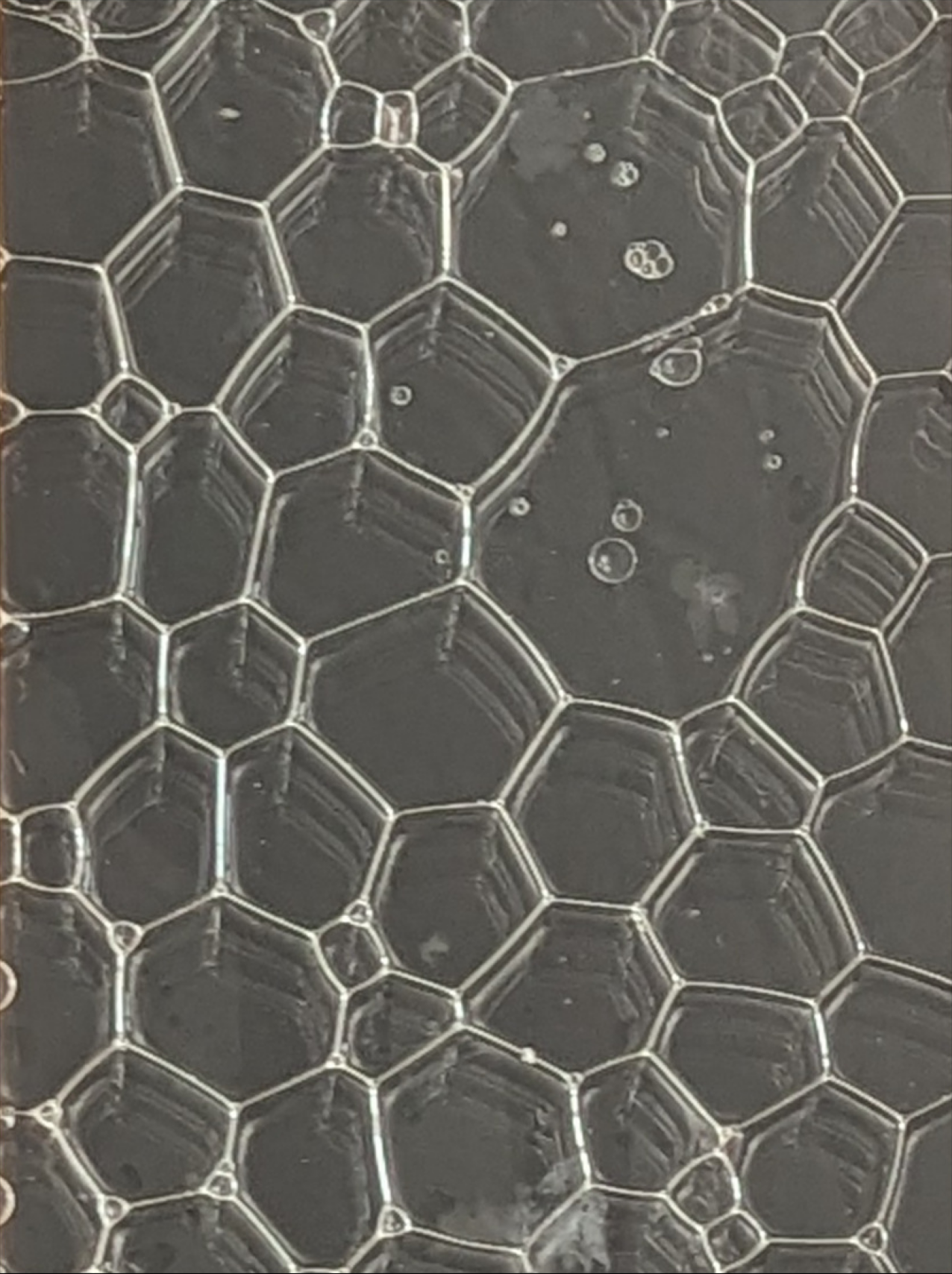}
  \caption{Input image.\vspace{3.52mm}}
\end{subfigure}
\begin{subfigure}{.32\linewidth}
  \centering
  \includegraphics[width=0.99\textwidth]{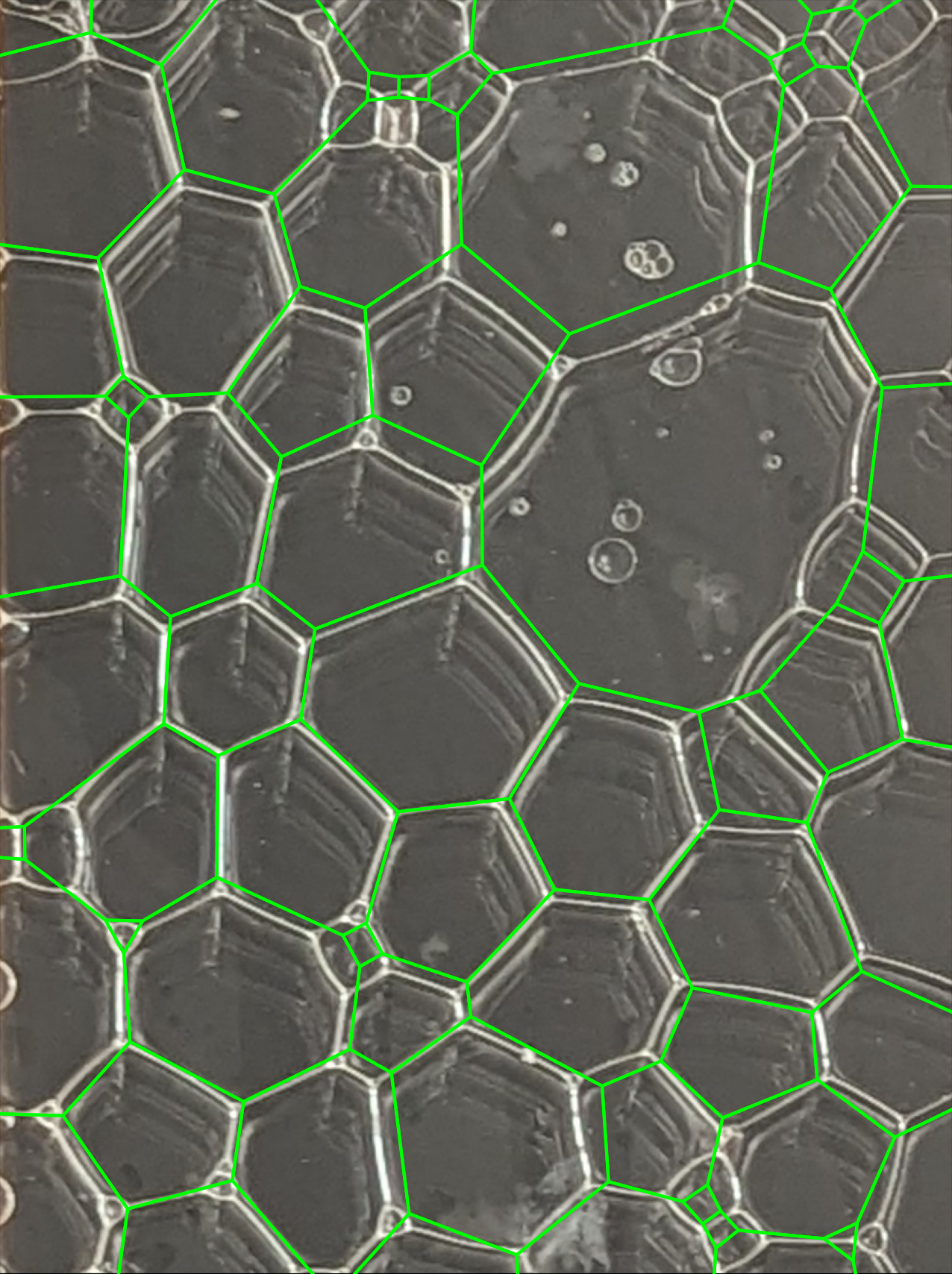}
  \caption{Before optimization. Objective value 0.28.}
\end{subfigure}
\begin{subfigure}{.32\linewidth}
  \centering
  \includegraphics[width=0.99\textwidth]{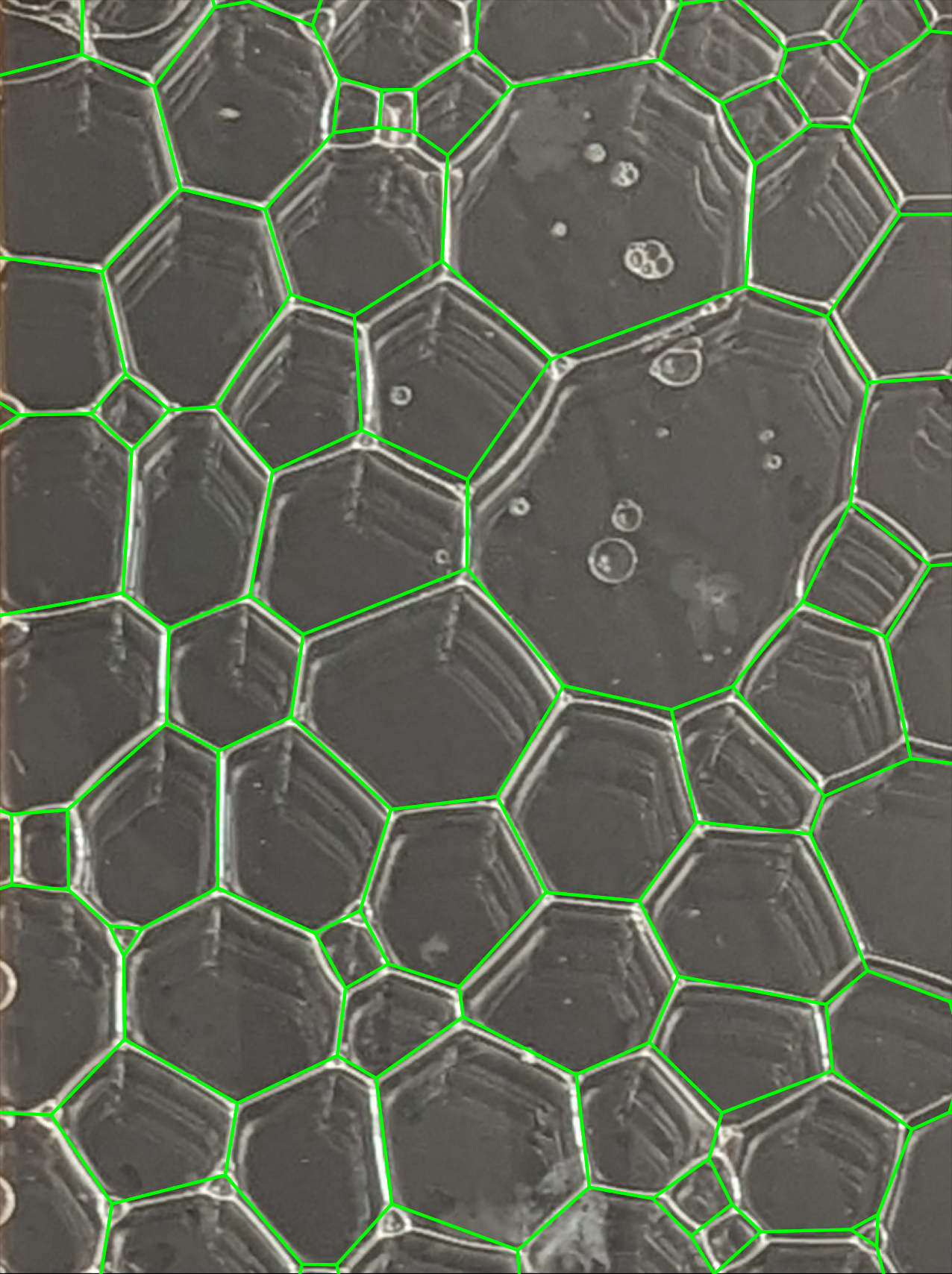}
  \caption{After optimization. Objective value 0.034.}
\end{subfigure}
\caption{Construction of a Voronoi foam model from an input image using equilibrium-constrained optimization.}
\label{fig:image_match}
\end{figure}

We apply our equilibrium-constrained optimization formulation (Sec.~\ref{sec:eq_constrained_optimization}) to characterize a 2D foam from an image and create a matching computational model. The image (Fig.~\ref{fig:image_match}\textit{a}) was taken in our lab and depicts a soap foam between two glass plates. We manually annotate the image, defining the vertex coordinates and the edges in the desired tessellation.
We assume a cell energy of the form
\begin{equation}\label{eq:image_match_energy}
    E = a_0(A-\bar{A})^2 + a_1P,
\end{equation}
where $A$ and $P$ are the cell area and perimeter, and $\bar{A}$ is a target area. This is analogous to the 3D dry foam energy used in our coarsening simulation (Sec.~\ref{sec:coarsening}).
The model is initialized by placing sites at the centroids of the cells in the annotated image, and assigning $\bar{A}$ equal to their area. The system is allowed to converge to equilibrium, resulting in the pre-optimization model shown in Fig.~\ref{fig:image_match}\textit{b}. 
We then perform equilibrium-constrained optimization on the area targets $\bar{A}$ to find cell pressures matching the observed configuration. The optimization objective is to minimize the sum of squared distances between the annotation vertices and the corresponding Voronoi vertices. The corresponding Voronoi vertex does not need to exist in the tessellation; the generating expression (the circumcenter of the three neighboring sites, or similar for boundary vertices) is always computable. 
%
The optimization converged in 77 L-BFGS iterations and 13 seconds of computation time.
As can be seen from Fig. ~\ref{fig:image_match}\textit{c}, the optimization successfully reduces the discrepancy between simulated and real-world foams.
It should be noted that this procedure converges reliably to a good solution only when the topology of the initial equilibrium state matches that in the image.

\subsection{Comparison to Deformable Cell Model}
To compare the accuracy and performance of our model with other simulation methods, we construct a deformable cell model (DCM) using Incremental Potential Contact (IPC)~\cite{Ipc2020} to resolve contact between the cells. The DCM cells have 30 vertices and hence 60 degrees of freedom, while ours are power diagram cells with 3 degrees of freedom. We use a per-cell energy of the form
\begin{equation}\label{eq:dcm_comparison_energy}
    E = a_0(A-\bar{A})^2 + a_1 P^2,
\end{equation}
where $A$ and $P$ are the cell area and perimeter, and $\bar{A}$ is a target area. We use a combined target of 1.2 times the domain area such that the DCM cells are in compression and fill the entire space. The DCM uses an additional quadratic penalty per edge length for regularization as well as the IPC energy.
\begin{figure}[h]
\centering
\begin{subfigure}{.38\linewidth}
  \centering
  \includegraphics[width=0.95\textwidth]{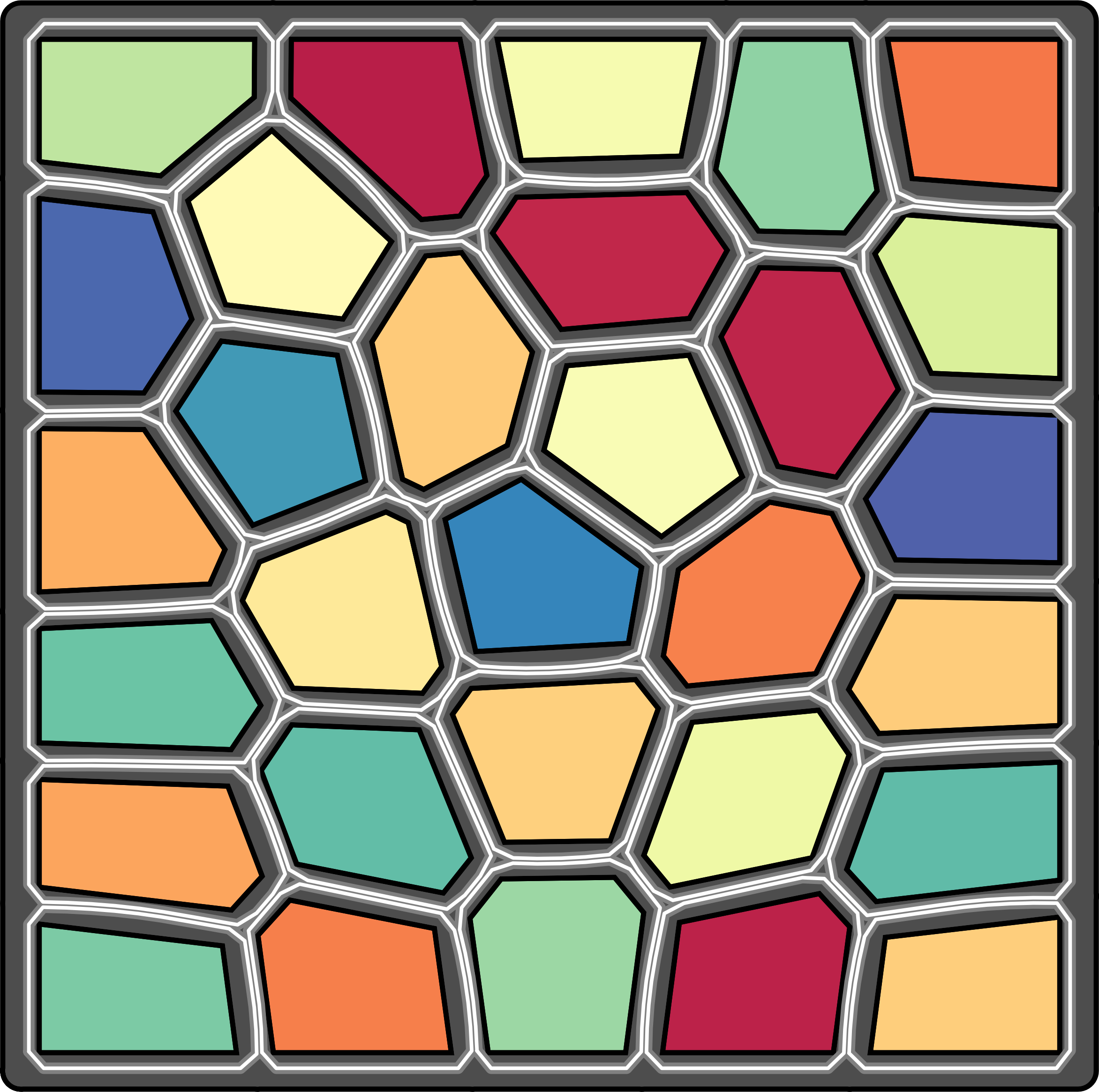}
  \caption{Initial state, 1x1 domain.}
\end{subfigure}
\begin{subfigure}{.55\linewidth}
  \centering
  \includegraphics[width=0.95\textwidth]{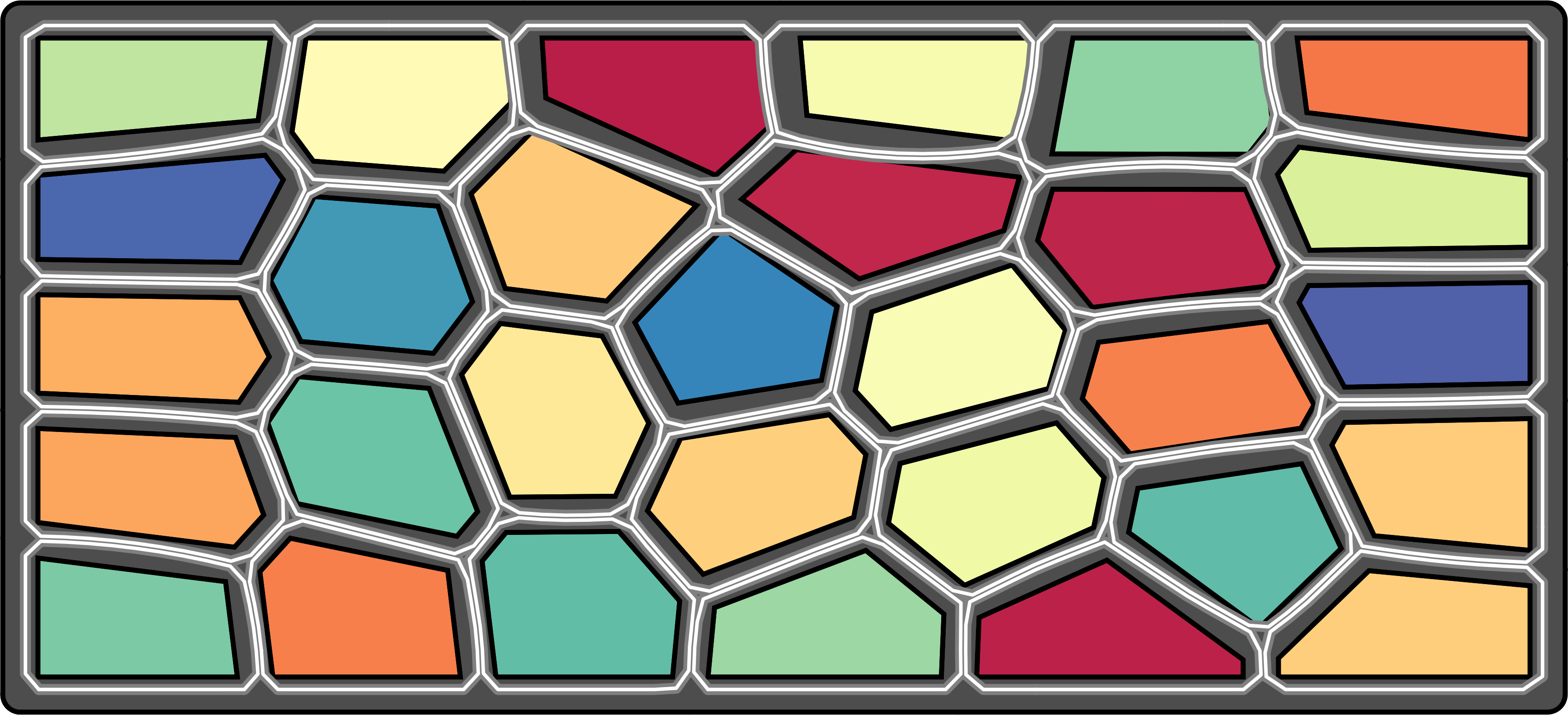}
  \caption{Final state, 1.50x0.67 domain.}
\end{subfigure}
\caption{Comparison between implicit Voronoi model (background) and deformable cell model (white overlay) in a 100-frame quasi-static simulation. Identical topology changes occur in both models.  Cells are colored to show correspondence between initial and final states.}
\label{fig:dcm_comparison}
\end{figure}
We perform a quasi-static simulation with 30 cells starting in a square domain. The initial state for the DCM is constructed by offsetting each Voronoi cell inward and distributing vertices evenly around the perimeter, followed by energy minimization to reach equilibrium. In each frame, the domain is reshaped and both models are allowed to converge to a new equilibrium. The two models behave similarly, achieving the same final topology as shown in Fig.~\ref{fig:dcm_comparison}. However, the DCM required 1000 times more computation time and nearly 40 times as many average Newton iterations to converge (see Table~\ref{tbl:timings}), demonstrating the efficiency of our implicit Voronoi model.
\subsection{Runtime}
Runtime statistics for all simulation experiments are collected in Table~\ref{tbl:timings}. Experiments are performed using a workstation with an AMD Ryzen Threadripper PRO 5995WX CPU.
\begin{table*}[ht]
    \caption{Runtime statistics for experiments. Each frame represents one simulation time step and consists of solving a single optimization problem.}
    \centering
    \begin{tabular}{p{3.5cm}rrrrrr}
    \toprule
       Experiment  & \# Cells & \# DOF & \# Frames & \# Iter / Frame (avg / max) & Time / Iter (avg) [ms] & Total time [s] \\
    \midrule
        Comparison (Ours) & 30 & 90 & 100 & 5/25 & 2.32 & 1.23 \\
        Comparison (DCM) & 30 & 1800 & 100 & 195/2621 & 65 & 1258 \\
    \midrule
        Tissue Growth (membrane) & 4096 & 24070 & 700 & 3/12 & 3830 & 8521 \\
        Tissue Growth (cylinder) & 620 & 6067 & 1101 & 10/47 & 1383 & 14970 \\
    \midrule
        Coarsening  & 2000 & 10000 & 350 & 51/522 & 1275 & 22996 \\
    \midrule
        Rigid Body 1 & 2000 & 4003 & 396 & 3/4 & 212 & 251 \\
        Rigid Body 2 & 2000 & 4003 & 304 & 3/5 & 213 & 194 \\
    \end{tabular}
    \label{tbl:timings}
\end{table*}
\par
The number of iterations per frame depends highly on the number and complexity of topological changes that occur, resulting in large discrepancies among the 3D experiments. In the embryonic cleavage example, few topological changes occur except in the frames immediately following each simultaneous cell division. The coarsening example is particularly expensive because many cells collapse per frame, and cell collapse results in more complex topological changes than intercalation. Furthermore, the energy gradient in the coarsening example is discontinuous across topology changes, slowing the convergence of Newton's method, while the other examples use $C^1$-continuous energies.

\begin{figure}[h]
\centering
\begin{subfigure}{.50\linewidth}
  \centering
  \includegraphics[width=0.9\textwidth]{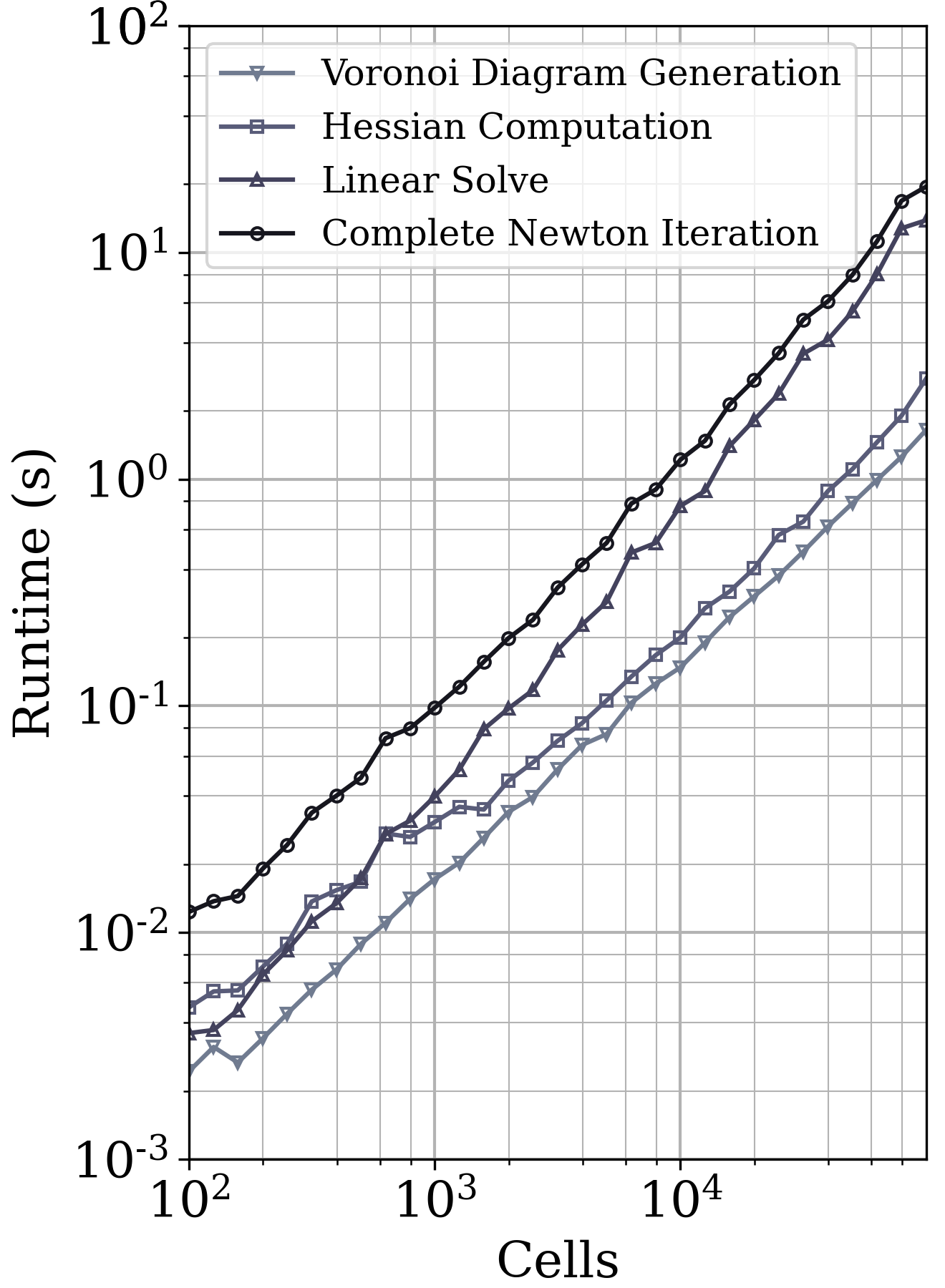}
\end{subfigure}
\begin{subfigure}{.48\linewidth}
  \centering
  \includegraphics[width=0.9\textwidth]{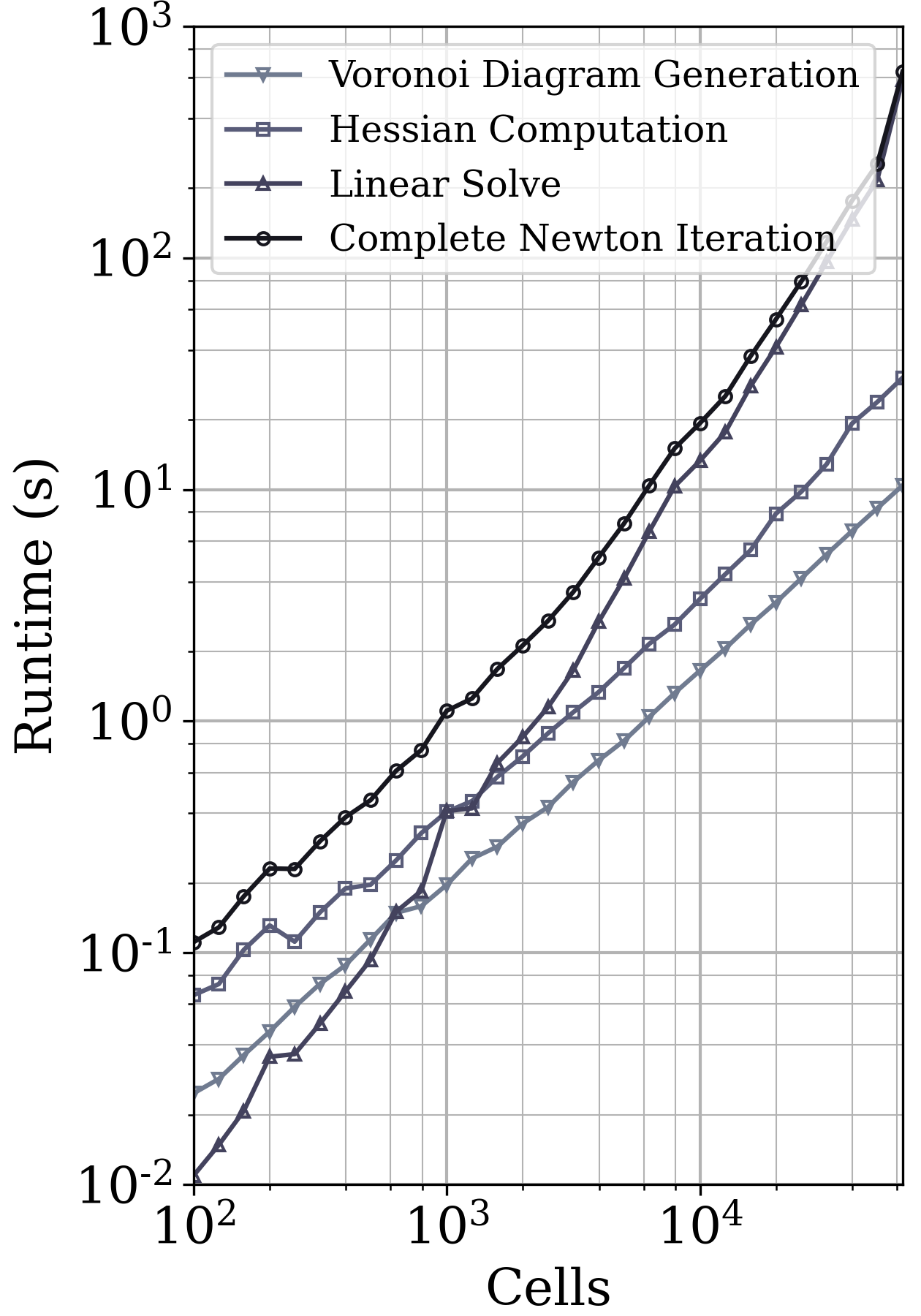}
\end{subfigure}
\caption{Runtime of a Newton iteration vs. number of cells for 2D (left) and 3D (right) models. In the range of problem sizes considered, the Voronoi diagram generation (including evaluation of derivatives e.g. $\frac{\dd \bx}{\dd \bc}$) and Hessian computation exhibit roughly linear scaling, while the cost of solving the linear system dominates for larger problems.}
\label{fig:runtime_scaling_plot}
\end{figure}

We perform an additional experiment to measure the runtime scaling of our method with number of cells. In a square (2D) or cube (3D) with $n$ randomly placed Voronoi sites, we perform a single Newton iteration and measure the total runtime, along with the runtime of significant subroutines including the generation of the Voronoi diagram, computation of the energy Hessian and solution of the linear system (Fig.~\ref{fig:runtime_scaling_plot}).
As can be observed from the results, 3D cases are significantly more expensive than 2D cases for similar numbers of cells. This is primarily due to the increase in average number of neighbors for each cell, from 6 in 2D to $\sim$15.5~\cite{Meijering1953} in 3D, leading to a much denser Hessian and a larger number of Voronoi vertices.

\section{Conclusions}

We have proposed a novel simulation approach for mechanical cellular systems based on differentiable Voronoi diagrams. The method successfully addresses challenges faced by existing models in handling topological transitions, while using an extremely compact state representation that implicitly defines the shape and topology of the interface network. We demonstrated this using a diverse set of examples, including simulations of complex biological and physical processes as well as comparisons to alternative models and real-world systems.

\subsection{Limitations \& Future Work}
The development of informative biological simulations will require more complex energy models and the incorporation of real-world data to determine accurate parameters. 
Deriving a smoother approximation to the surface area of Voronoi cells would enable further applications, including adhesion effects at biological cell interfaces.
Our model currently uses an isotropic distance metric that restricts the space of possible cell shapes, and the method assumes planar or piecewise linear cell interfaces. However, epithelial tissues, for example, exhibit non-convex scutoid-shaped cells~\cite{gomez2018scutoids} which permit different cell packing patterns. Extending our method towards more complex cell shapes while retaining a compact site-based representation is an interesting challenge for future work.
Voronoi diagrams are also a natural design space for foam-like 3D-printed metamaterials. This direction has been explored, \eg, by Martinez \etal~\shortcite{Martinez2018}, with extensions
to orthotropic foams based on Voronoi diagrams with spatially varying distance metrics~\cite{StarShapes}. An extension of our approach may permit gradient-based optimization of these materials.
Finally, we have investigated a simple instance of intercellular navigation, where a rigid object is driven by a constant force. Designing control algorithms to guide an object through cellular systems towards a desired target is another exciting future direction.
\begin{acks}
We thank the anonymous reviewers for their valuable feedback. This work was supported by the European Research Council (ERC) under the European Union’s Horizon 2020 research and innovation program (grant agreement No. 866480), and the Swiss National Science Foundation through SNF project grant 200021\_200644.
\end{acks}
\bibliographystyle{ACM-Reference-Format}
\bibliography{reference}
\appendix

\section{Volume integration in tetrahedra}\label{sec:tetrahedron_integration}
Here we provide a typical approach for volume integration over a tetrahedron, as used for analytically evaluating certain cell energy functions in this work. We consider a tetrahedron with vertices ($x_i$, $y_i$, $z_i$), $i=0,1,2$ and the origin.
The tetrahedron can be parameterized with coordinates ($u$, $v$, $w$) $\in$ [0,1] such that
\begin{equation}\label{eq:tet_parameterization}
\begin{aligned}
x = u(vx_0+(1-v)(wx_1+(1-w)x_2)), \\
y = u(vy_0+(1-v)(wy_1+(1-w)y_2)), \\
z = u(vz_0+(1-v)(wz_1+(1-w)z_2)).
\end{aligned}
\end{equation}
The Jacobian matrix of this transformation is
\begin{equation}\label{eq:tet_parameterization_jacobian}
J = \begin{bmatrix} \frac{\partial x}{\partial u} & \frac{\partial x}{\partial v} & \frac{\partial x}{\partial w} \\ \frac{\partial y}{\partial u} & \frac{\partial y}{\partial v} & \frac{\partial y}{\partial w} \\ \frac{\partial z}{\partial u} & \frac{\partial z}{\partial v} & \frac{\partial z}{\partial w} \end{bmatrix} 
\end{equation}
where the partial derivatives are
\begin{equation}\label{eq:tet_parameterization_jacobian_terms}
\begin{aligned}
\frac{\partial x}{\partial u} = & vx_0+(1-v)(wx_1+(1-w)x_2), \\
\frac{\partial x}{\partial v} = & u(x_0-wx_1-(1-w)x_2), \\
\frac{\partial x}{\partial w} = & u(1-v)(x_1-x_2).
\end{aligned}
\end{equation}
A volume integral can be transformed as follows, where $f^*$ is obtained by substituting equations~\ref{eq:tet_parameterization} for ($x, y, z$) in $f$:
\begin{equation}
\iiint f(x,y,z) dV = \int \limits_0^1 \!\!\!\! \int \limits_0^1 \!\!\!\! \int \limits_0^1 \det(J) \hspace{1mm} f^*(u,v,w) \hspace{1mm} du dv dw
\end{equation}

\section{Hessian for Boundary Coupling}
\label{sec:appHessianBoundary}
Taking the derivative of (\ref{eq:energy_grad_with_boundary}), we obtain the second order terms 
\begin{equation}\label{eq:energy_hess_with_boundary}
\begin{gathered}
\begin{aligned}
\frac{\dd^2 E}{\dd \bc \dd \bp}
& = \left(\frac{\partial \bx}{\partial \bc}\right)^\top\frac{\partial^2 F}{\partial \bx^2}\frac{\partial \bx}{\partial \bv}\frac{\dd \bv}{\dd \bp} + \left(\frac{\partial \bx}{\partial \bc}\right)^\top\frac{\partial^2 F}{\partial \bx \partial \bp} \\
& + \frac{\partial^2 F}{\partial \bc \partial \bx}\frac{\partial \bx}{\partial \bv}\frac{\dd \bv}{\dd \bp} + \left(\sum_i\frac{\partial F}{\partial \bx_i}\frac{\partial^2 \bx_i}{\partial \bc \partial \bv}\right)\frac{\dd \bv}{\dd \bp} + \frac{\partial^2 F}{\partial \bc \partial \bp},
\end{aligned} \\
\begin{aligned}
\frac{\dd^2 E}{\dd \bp^2}
& = \left(\frac{\partial \bx}{\partial \bv}\frac{\dd \bv}{\dd \bp}\right)^\top \frac{\partial^2 F}{\partial \bx^2} \frac{\partial \bx}{\partial \bv}\frac{\dd \bv}{\dd \bp} + \left(\frac{\partial \bx}{\partial \bv}\frac{\dd \bv}{\dd \bp}\right)^\top \frac{\partial^2 F}{\partial \bx \partial \bp} \\
& + \frac{\partial^2 F}{\partial \bp \partial \bx} \frac{\partial \bx}{\partial \bv}\frac{\dd \bv}{\dd \bp} + \left(\frac{\dd \bv}{\dd \bp}\right)^\top\left(\sum_i\frac{\partial F}{\partial \bx_i}\frac{\partial^2 \bx_i}{\partial \bv^2}\right)\frac{\dd \bv}{\dd \bp} \\
& + \left(\sum_i\left(\frac{\partial F}{\partial \bx}\frac{\partial \bx}{\partial \bv_i}\right)\frac{\dd^2 \bv_i}{\dd \bp^2}\right) + \frac{\dd^2 F_B}{\dd \bp^2} \ ,
\end{aligned} \\
\end{gathered}
\end{equation}
which must be added to the Hessian (\ref{eq:energy_hess_no_boundary}) of the fixed-boundary energy.

\section{Dynamic simulation convergence analysis}\label{sec:convergence_analysis}
\begin{figure}[h]
\centering
\begin{subfigure}{.48\linewidth}
  \centering
  \includegraphics[width=0.75\textwidth]{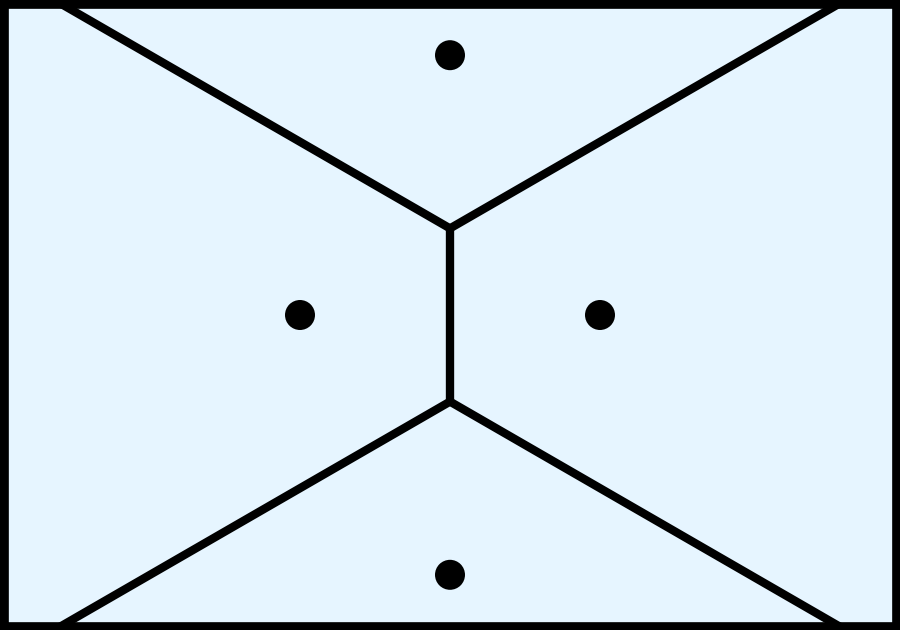}
\end{subfigure}
\begin{subfigure}{.48\linewidth}
  \centering
  \includegraphics[width=0.75\textwidth]{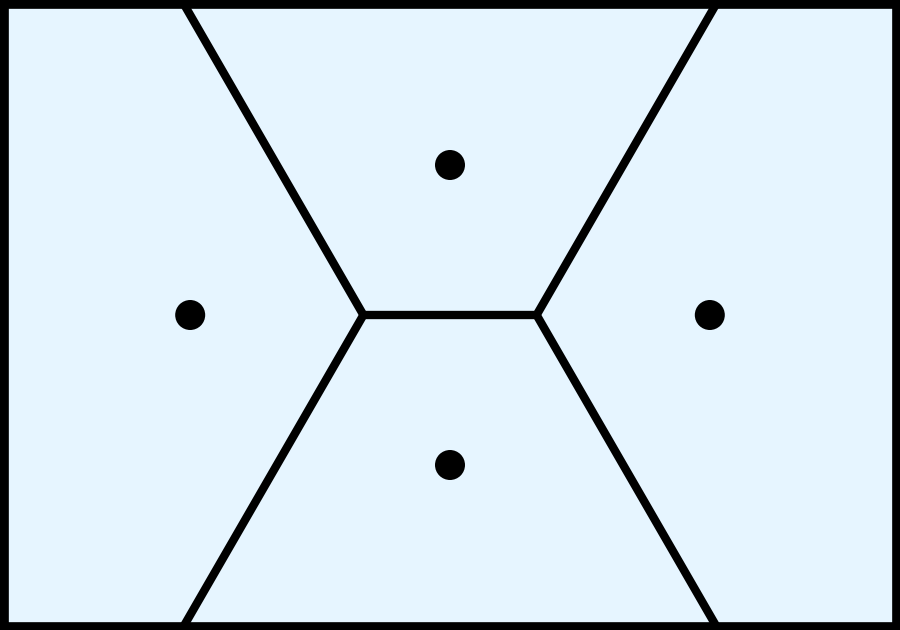}
\end{subfigure}
\caption{Approximate initial (left) and final (right) states of a dynamic simulation used to assess the convergence of the implicit Voronoi model.}
\label{fig:convergence_test_scenario}
\end{figure}

\begin{figure}[H]
  \centering
  \includegraphics[width=0.9\linewidth]{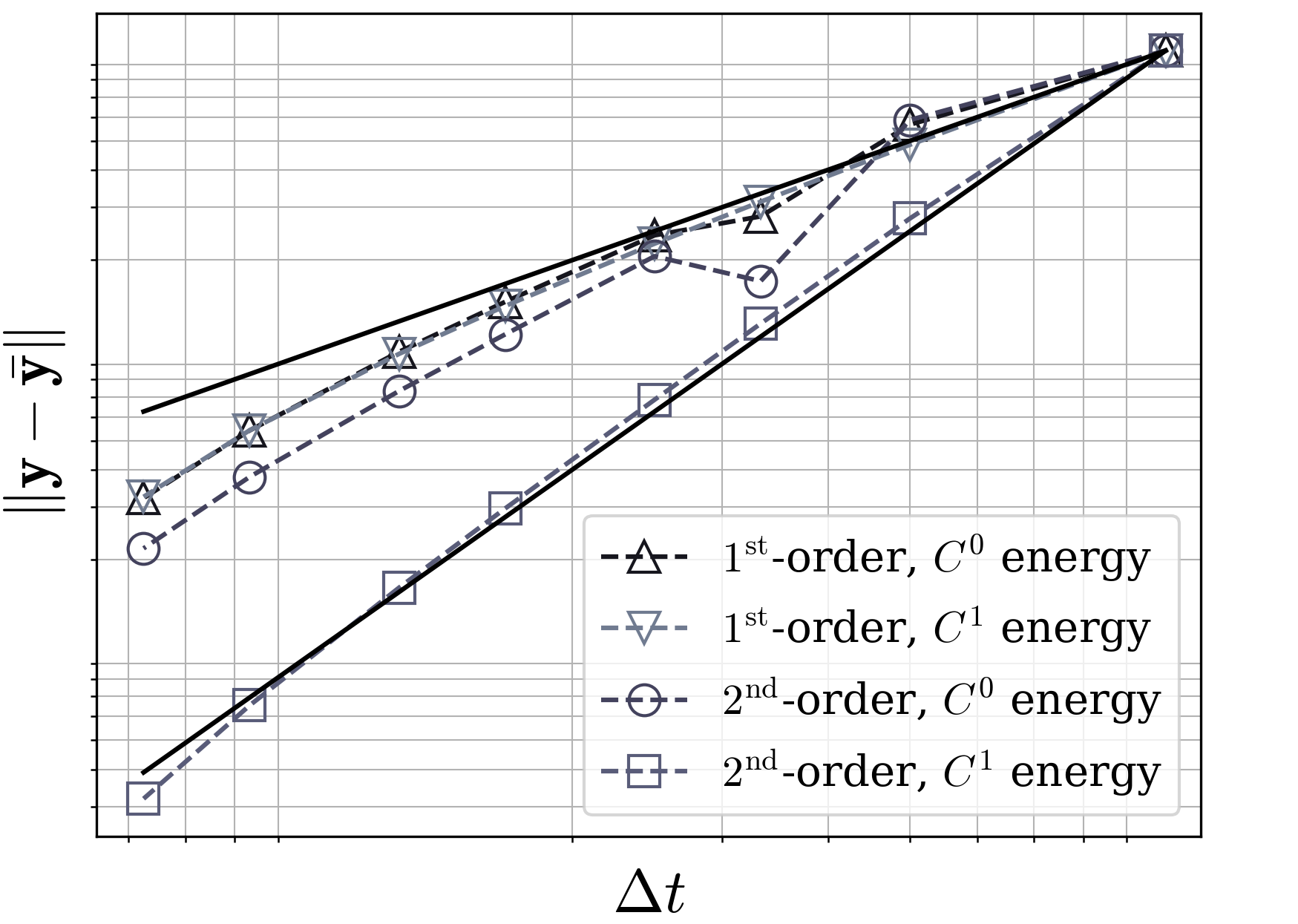}
  \caption{Final state error vs. time step for a simple 2D case including a topological transition. Final state from highest resolution simulation is taken as ground truth $\bar{\by}$. Solid lines represent exact linear and quadratic convergence, with steeper slope indicating higher-order convergence.}
  \label{fig:convergence_test_plot}
\end{figure}

Convergence of simulations under refinement of the time step is important for assessing the accuracy of numerical solutions. To evaluate the convergence of dynamic simulations using our method, we perform a simple 2D simulation with four cells and one topological transition, illustrated in Fig.~\ref{fig:convergence_test_scenario}.
The simulation is repeated for the same total time using a range of time steps. Error measurements in Fig.~\ref{fig:convergence_test_plot} show that given a sufficiently smooth energy function and using second-order finite difference approximations (BDF2), the dynamic simulation method presented in section~\ref{sec:dynamics} converges quadratically. Linear convergence is achieved for $C^0$-continuous energy functions (such as perimeter and surface area) or using only a first-order accurate acceleration.
%
\section{Experiment Details}\label{sec:experiment_details}
In this section we provide additional details and parameters required to reproduce the examples in Sec.~\ref{sec:results}.
\paragraph{Intercellular Navigation with Rigid Body}
The domain is rectangular with width 4\unit{m} and height 2\unit{m}, containing $n=2000$ cells along with the rigid body. Other parameters are $\bar{A}=0.004$~\unit{m^2}, $a_0 = 0.05$~\unit{N/m^3}, $a_1 = 100.0$~\unit{N/m^3}, $a_2 = 1.0$~\unit{N/m}, $F_x=0.035$~\unit{N}, $h = 0.01$~\unit{s}, $m_\bc=0.0003$~\unit{kg}, $m_\bp=0.0003$~\unit{kg}, $\eta_\bc=0.0003$~\unit{N.s/m}, $\eta_\bp=0.003$~\unit{N.s/m}. Here and in the following examples, $m$ and $\eta$ are diagonal entries of the mass matrix $\mathbf{M}$ and viscosity matrix $\boldsymbol{\eta}$ corresponding to the site ($\bc$) and boundary ($\bp$) degrees of freedom.
\paragraph{Foam Coarsening}
The domain is a 2\unit{m}$\times$2\unit{m}$\times$20\unit{cm} box starting with $n=2000$ cells. Other parameters are $\bar{V}_\text{initial}=0.0004$~\unit{m^3}, $a_0 = 0.002$~\unit{N.m}, $a_1 = 0.01$~\unit{N/m}, $a_2 = 0.1$~\unit{N/m}, $\eta_\bc=0.0001$~\unit{N.s/m}, $\eta_{\bar{V}}=1.0$~\unit{N.s/m^3}. The time step $h=\frac{1.0}{n}$~\unit{s} is adapted each frame to the number of remaining cells $n$, increasing as the rate of cell collapse slows.
\paragraph{Embryonic Cleavage}
The initial domain is an \textit{icosphere} with 4 subdivisions, approximating a sphere with radius 1.34\unit{m}. Other parameters are $\bar{V}_\text{initial}=10.25$~\unit{m^3}, $a_0 = 50.0$~\unit{N.m}, $a_1 = 5.0$~\unit{N}, $a_2 = 1.0$~\unit{N.m}, $h = 0.01$~\unit{s}, $\eta_\bc=5.0$~\unit{N.s/m}, $\eta_{\bp}=0.5$~\unit{N.s/m}, $\beta = 0.1$.
\paragraph{Tissue Proliferation in Cylinder}
The initial domain is a cylinder of radius 1\unit{m} whose upper deformable face comprises 2432 triangles. Other parameters are $\bar{V}_\text{initial}=0.113$~\unit{m^3}, $a_0 = 10.0$~\unit{N.m}, $a_1 = 10.0$~\unit{N}, $a_2 = 1.0$~\unit{N.m}, $h = 0.01$~\unit{s}, $\eta_\bc=0.5$~\unit{N.s/m}, $\eta_{\bp}=0.003$~\unit{N.s/m}, $\alpha = 30$~\unit{m^{-3}.s^{-1}}, $\beta = 0.1$, $\gamma = 0.75$, $\tau = 0.11$~\unit{s}.

\end{document}